\documentclass[%
 preprint,
 amsmath,amssymb,
 aps,
 showkeys,
]{revtex4-2}

\usepackage{graphicx}% Include figure files
\usepackage{dcolumn}% Align table columns on decimal point
\usepackage{bm}% bold math
\usepackage{hyperref}% add hypertext capabilities
\usepackage{xcolor}
\usepackage{float}

\newcommand{\diff}{\mathrm{d}}

\begin{document}

\title{Light Curves of the Hot Spots on Circular Orbits Around Generic Regular Black Holes Related to Non-linear Electrodynamics with Maxwellian Weak-field Limit}

\author{Dmitriy Ovchinnikov}
\email{dmitriy.ovchinnikov@physics.slu.cz} 
\affiliation{Research Centre for Theoretical Physics and Astrophysics, Institute of Physics, Silesian University in Opava, Bezrucovo namesti 13, CZ-746 01 Opava, Czech Republic}

\author{Jan Schee}
\email{jan.schee@physics.slu.cz}
\affiliation{Research Centre for Theoretical Physics and Astrophysics, Institute of Physics, Silesian University in Opava, Bezrucovo namesti 13, CZ-746 01 Opava, Czech Republic}

\author{Zdenek Stuchl{\'\i}k}
\email{zdenek.stuchlik@physics.slu.cz} 
\affiliation{Research Centre for Theoretical Physics and Astrophysics, Institute of Physics, Silesian University in Opava, Bezrucovo namesti 13, CZ-746 01 Opava, Czech Republic}

\author{Bobomurat Ahmedov}
\email{ahmedov@astrin.uz} 
\affiliation{Ulugh Beg Astronomical Institute, Astronomy St. 33, Tashkent 100052, Uzbekistan}
\affiliation{Tashkent Institute of Irrigation and Agricultural Mechanization Engineers, Kori Niyoziy, 39, Tashkent 100000, Uzbekistan}
\affiliation{Physics Faculty, National University of Uzbekistan, Tashkent 100174, Uzbekistan}

\date{\today}

\begin{abstract}
The combined effect of a strong gravitational field and electromagnetic field in the vicinity of a generic regular black hole related to non-linear electrodynamics with Maxwellian weak-field limit on the radiation flux of a hot spot orbiting the regular black hole on the Keplerian disc has been studied. The frequency shift due to the strong gravitational field and magnification of the radiation related to gravitational lensing have been calculated. We compare the flux related to the Maxwellian regular black hole to the flux of the hot spots moving with the same orbital period around standard Schwarzschild and Raissner-Nordstrom black holes to illustrate the role of the effective geometry governing photon motion around the regular black hole.   
\end{abstract}


\maketitle

%\tableofcontents

\section{\label{sec:intro}Introduction}

One of the possible mechanisms explaining the variable brightness from X-ray sources in the center of galaxies may be associated with inhomogeneities arising on the inner part of the accretion discs, from where the main part of the X-ray radiation comes ~\cite{abramowicz1991x}. The size, different intensities, and time inconsistency of such areas may essentially change the radiation flux ~\cite{bao1992signature, stuchlik1992radiation} i.e., light curve, leading to the time variability of X-ray sources.

Another crucial optical effect giving relevant observational predictions is the shape of profiled spectral lines (usually Fe lines) generated by inner parts of the Keplerian discs or slender torus ~\cite{laor1991line, bao1992accretion, schee2009profiles}. The profiled spectral lines were intensively studied ~\cite{schee2013profiled, schee2019profiled, schee2016profiled, bambi2013testing, bambi2013k} as they enable relatively precise estimates of the black hole spin ~\cite{shafee2005estimating, mcclintock2011measuring, steiner2011spin}. In the present paper, we concentrate attention on study of the hot spot light curves, as they could be a promising model for describing optical phenomena on supermassive black hole  Sgr A* representing an underdense nucleus of the Milky Way Galaxy.

When studying the processes occurring in the close vicinity of black holes, it is necessary to take into account the relativistic effects associated with a strong gravitational field ~\cite{zhang1991rotation} and ultrarelativistic velocities. Corresponding adjustments to the radiation characteristics are usually introduced by taking into account gravitational redshift and gravitational lensing ~\cite{cunningham1973optical, cunningham1975effects, bao1992signature}. Furthermore, since the inhomogeneities on the inner part of the accretion discs move with the accretion disc, the vector of its velocity is constantly changed, leading to the special relativistic Doppler effect and associated Doppler beaming. Let us stress that for the relativistic Keplerian orbit near the black hole horizon also the time delay effects plays an important role in modeling the light curves of the hot spot.

As a source of modeling radiation from irregularities on the inner part of the accretion discs, one can assume that a group of small but finite size radiating spheres or spots are rotating together with the accretion disc in Keplerian orbits ~\cite{bao1992signature}.

As a model of a black hole, we consider generic regular black holes related to non-linear electrodynamics with Maxwellian weak-field limit (Maxwellian regular black hole for short). Construction of the generic regular black holes is discussed and studied in ~\cite{fan2016construction, toshmatov2018comment, abdujabbarov2017gravitational, toshmatov2018electromagnetic, toshmatov2019relaxations}. The main interest to regular solutions for black holes is related to the fact that such solutions do not contain the central physical singularities where the geometry and curvature of spacetime diverge. The notion of regular black holes was first introduced by Bardeen in ~\cite{bardeen1968non}. Furthermore more general exact solutions for the regular black hole were  obtained in ~\cite{ayon1998regular, ayon1999new} by solving the Einstein field  equations, assuming that the action contains terms that admit nonlinearity of the coupled  electromagnetic field equations. Relation of such solutions to non-linear electrodynamics was extensively explored in ~\cite{bronnikov2001regular, bronnikov2017comment, bronnikov2000comment}. The rotating regular black holes were in the framework of the non-linear electrodynamics introduced and studied in a variety of papers ~\cite{bambi2013rotating, azreg2014generating, toshmatov2014rotating, toshmatov2017rotating}. In the present paper, we restricted attention to the spherically symmetric regular black holes, since such a restriction enables an efficient description of the role of the direct electromagnetic effect governed by the effective geometry ~\cite{stuchlik2019shadow}.

The peculiarities in studying the motion of particles in the vicinity of the Maxwellian regular black hole is that the motion of charged particles and photons is due to the influence of non-linear electrodynamics should be considered in effective geometry ~\cite{stuchlik2019shadow, stuchlik2019generic}. The effective geometry of spacetime  takes into account the direct effects of the non-linear electrodynamics, along with the effect reflected in the spacetime geometry. We are thus extending our previous studies related to the direct influence of the non-linear electromagnetic fields on the optical phenomena due to the effective geometry ~\cite{stuchlik2019shadow, stuchlik2015circular, schee2015gravitational,abdujabbarov2017gravitational}.

In the present work we study the influence of strong gravitational field  in the vicinity of Maxwellian regular black hole and its non-linear electromagnetic field on the radiation flux from the hot spots moving on the innermost stable circular orbits (ISCOs) around regular  black holes. We have calculated the frequency shift and magnification of the radiation caused by the gravitational lensing. In order to analyze fingerprints of Maxwellian regular black hole we have constructed the light curve of the hot spots moving around Maxwellian regular black hole and compared it with the light curve of the hot spots moving with the same orbital period  around Schwarzschild ~\cite{bao1992signature} and Reissner-Nordstrom ~\cite{stuchlik1992radiation} black holes.    

The paper is organized as follows. Section~\ref{sec:hot-spot-rad} is devoted to detailed description of hot spots radiation. The next Section~\ref{sec:reg-bh} describes spacetime of the regular black hole with the magnetic charge coupled to NED. Section~\ref{sec:eq-of-motion} is devoted to the equations of motion  of photons and neutrinos especially the motion of photons  determined by geodesics in the effective geometry, including both the effects of spacetime geometry and the effects of NED. Light curves construction of radiation from a small hot spot in a Keplerian circular orbit to study the effects of a strong gravitational field in the vicinity of a regular black hole related to NED is performed in Section~\ref{sec:light-curves}. In the Section~\ref{sec:results} we have calculated the frequency shift, gravitational lensing and have constructed the light curves of the hot spots on ISCOs and the hot spots with equal orbital periods for different values of the charge parameter $q_m$ moving around Maxwellian regular black hole.  Also, we have compared the results  obtained for the hot spots radiation to that moving with the equal orbital periods around Schwarzschild and Reissner-Nordstrom black holes. Discussion and  analysis of possible  fingerprints of Maxwellian regular black hole in light curve of point-like hot spot orbiting on Keplerian circular orbit is summarized in the last 
Section~\ref{sec:dis-concl}.

\section{\label{sec:hot-spot-rad}Radiation of Hot Spot orbiting black hole}

Consider a small spherical radiating source in a circular Keplerian orbit in the vicinity of a black hole assuming that the radiation from the source has the properties of blackbody radiation. The total radiation flux $F_o$ from such source to a distant observer is determined by the radiation intensity at the observation point $I_o$ and the solid angle $\diff\Omega$ at which this source is visible. Integrating by all of the frequencies $\nu_o$ at the point of observation, the equation for the total flux $F_o$ reads as
\begin{equation}
    F_o = I_o \, \diff \Omega\ .
    \label{eq:flux}
\end{equation}
According to Liouville's theorem, when some number of identical particles $\delta N$ move in the spacetime, the corresponding volume in phase space $V \equiv V_x V_p $ is conserved, and accordingly, the ``number density in phase space'' $\mathcal{N}$, for such group of particles remains unchanged for different Lorentz observers: 
\begin{equation}
    \mathcal{N} \equiv \frac{\delta N}{V} = \frac{\delta N}{h^3 \, \nu^2 \, \Delta A \, \Delta t \, \Delta\nu \, \Delta\Omega}\ ,
    \label{eq:num-dens}
\end{equation}
where $h$ is the Planck constant. 

However, in astronomy, to describe the energy of radiation, instead of considering the energy of individual photons, it is more convenient to use the specific intensity $I_{\nu}$
\begin{equation}
    I_{\nu} \equiv \frac{\diff E_{\nu}}{\diff A \, \diff t \, \diff \nu \, \diff \Omega}\ .
    \label{eq:spec-intens}
\end{equation}

The introduced quantity~(\ref{eq:spec-intens}) describes the amount of radiation energy $\diff E_{\nu}$ of a given frequency $\diff \nu$ coming to the area $\diff A$, inside the solid angle $\diff \Omega$, over time $\diff t$. Comparing the expression for the intensity ~(\ref{eq:spec-intens}) with the expression for the ``number density in phase volume'' ~(\ref{eq:num-dens}), and using the relation $\diff E_{\nu} = h \nu \, \delta N$, it can be shown that for different Lorentz observers the value $I_{\nu}/\nu^3$ is also conserved. Therefore we obtain the fundamental expression relating the intensity and frequency of radiation at the place of emission and observation

\begin{equation}
    \frac{I_{\nu o}}{\nu_o^3} = \frac{I_{\nu e}}{\nu_e^3} = const.
    \label{eq:intens-freq-rel}
\end{equation}

The change in the frequency of radiation is caused by the influence of the gravitational redshift and Doppler effect in the case of a non-zero component of the emitting object's velocity on the direction of the line of sight. To describe the shift in radiation frequency, for convenience, the factor $g$ is introduced. This factor is equal to the ratio of the frequency of radiation in the observer's reference frame $\nu_o$ to the frequency of radiation in the reference frame associated with the source $\nu_e$
\begin{equation}
    g \equiv \frac{\nu_o}{\nu_e}.
    \label{eq:red-factor}
\end{equation}

Substituting the frequency shift coefficient $g$ in equation ~(\ref{eq:intens-freq-rel}), we obtain the relation for the specific intensity at the place of radiation $I_{\nu e}$ and at the place of observation $I_{\nu o}$
\begin{equation}
    I_{\nu o} = g^3 I_{\nu e}.
    \label{eq:spec-intens-obs-em-rel}
\end{equation}

By integrating the last expression over all possible frequencies in the radiation spectrum of the studying object one can get 
\begin{equation}
    I_o = \int_{0}^{\infty} I_{\nu o} \, \diff\nu_o = \int_{0}^{\infty} I_{\nu e} \, g^3 \, \diff\nu_o = 
    \int_{0}^{\infty} I_{\nu e} \, g^3 \, \diff(g \, \nu_e) = g^4 \int_{0}^{\infty} I_{\nu e} \, \diff\nu_e\ , \nonumber
\end{equation}
and obtain the well known expression ~\cite{misner1973gravitation} relating the total intensity at the place of emission $I_e$ with the total intensity at the place of observation $I_o$
\begin{equation}
    I_{o} = g^4 I_{e}.
    \label{eq:intens-obs-em-rel}
\end{equation}

\section{\label{sec:reg-bh}Spacetime of regular black hole related to NED}

One of the frequently used methods for obtaining solutions for regular black holes is to modify a standard Lagrangian density function that would include terms associated with NED. The action integral for such a function can be written as
\begin{equation}
    S = \frac{1}{16\pi}\int\diff x^4\sqrt{-\det(g_{\mu\nu})}\left[R-\mathcal{L}(F)\right]\ , 
    \label{eq:action}
\end{equation}
where $g_{\mu\nu}$ is the metric tensor, $R$ represents its Ricci scalar, $\mathcal{L}$ represents the NED (Non-linear Electrodynamics) Lagrangian density function, and the Faraday scalar $F\equiv F_{\alpha\beta}F^{\alpha\beta}$ is constructed from the electromagnetic field tensor $F_{\alpha\beta}=\partial_\alpha A_\beta - \partial_\beta A_\alpha$ related to the vector potential $A^{\alpha}$. There are different kinds of NED Lagrangian density functions, which allow regular black hole solutions do not containing spacetime singularities ~\cite{dymnikova2004regular, bronnikov2001regular, balart2014regular}. 

In this paper, we consider the static and spherically symmetric solutions with the line element taking in the standard Schwarzschild coordinates and geometric units ($c=G=1$) the form
\begin{equation}
    \diff s^2 = -f(r)\diff t^2 + \frac{1}{f(r)}\diff r^2 + r^2\diff\theta^2 + r^2\sin^2\theta\diff\phi^2,
    \label{eq:geometry}
\end{equation}
with the lapse function $f(r)$. Then the electromagnetic vector potential can be written in general form 
\begin{equation}
    A_{\alpha} = \Psi(r)\delta^{t}_{\alpha} - q_{m} cos\theta \delta^{\phi}_{\alpha} , 
\end{equation}
where $\Psi(r)$ denotes the electric potential function, and $q_m$ denotes the magnetic charge. The generic electrically or magnetically charged NED spacetimes were introduced in \cite{fan2016construction} and discussed in \cite{bronnikov2017comment,toshmatov2018comment}. 

The non-linear electromagnetic Lagrangian takes the generic form \cite{fan2016construction}
\begin{equation}
	\mathcal{L}(F)=\frac{4\mu}{\alpha}\frac{(\alpha F )^{\frac{\nu+3}{4}}}{[1+(\alpha F)]^{1+\frac{\mu}{\nu}}}\ ,
\end{equation}
where $\alpha$ represents an intensity parameter related to the charge  (details on the relation to the charge and mass parameters can be found in \cite{fan2016construction, toshmatov2018comment}), and Faraday scalar $F$ in case ($\Psi=0, q_m \neq 0$) takes the form
\begin{equation}
	F=\frac{1}{\alpha}\frac{q_m^4}{r^4}\ .
\end{equation} 
The parameters $\mu \geq 0$ and $\nu > 0$ are dimensionless constants of the spacetime, the parameter $\mu$ characterizes the electrodynamic non-linearity.

For the regular black hole with the magnetic charge related to NED, the lapse function takes the form \cite{fan2016construction, toshmatov2018comment} 
\begin{equation}
	f(r)=1-\frac{2M}{r}+\frac{2q_{m}^{3}}{\alpha r}-\frac{2q_{m}^{3}}{\alpha}\frac{ r^{\mu-1}}{(r^{\nu}+q_{m}^{\nu})^{\frac{\mu}{\nu}}}\ .
	\label{eq:lapse_1}
\end{equation}

This solution is parametrized by the mass parameter $M$ and the magnetic charge parameter $q_m$. As discussed in \cite{toshmatov2018comment} the only way to make metric regular everywhere in the spacetime is to assume that gravitational mass is equal to $M=q_m^3/\alpha$ and restrict ourselves to considering the values of the parameter $\mu \geq 3$.

The light curves obtained for the Maxwellian regular black hole  will be  compared with appropriate hypothetical graphs of neutrino flux over a period of time which could be obtained by the distant observer. We will also compare the light curves obtained for the Maxwellian regular black hole with the light curves obtained for the Reissner-Nordstrom black hole, which has the magnetic charge equal to $q_m$, and with the limiting case of a vanishing charge when the solution for a charged black hole turns into a solution for a Schwarzschild one.  

In the Reissner-Nordstrom spacetime for the lapse function ~(\ref{eq:geometry}) in the line element ~(\ref{eq:lapse_1}) one can use
\begin{equation}
	f(r) = 1-\frac{2M}{r}+\frac{q_m^2}{r^2},
	\label{eq:lapse_2}
\end{equation}
where parameter $q_m$ is the magnetic charge, and $M$ is the mass of the central black hole. For the Schwarzschild spacetime, the charge parameter $q_m = 0$, and the lapse function ~(\ref{eq:lapse_2}) reduces to
\begin{equation}
	f(r) = 1-\frac{2M}{r}.
	\label{eq:lapse_3}
\end{equation}

\section{\label{sec:eq-of-motion}Equations of motion of test particles}

Photons and neutrinos moving in the Schwarzschild or Reissner-Nordstrom spacetime do not affect the spacetime geometry and move along the null geodesics of spacetime geometry. However, when there is the influence of NED effects, while the spacetime geometry still governs the motion of neutrinos, the motion of photons is determined by geodesics in the effective geometry, including both the effects of spacetime geometry and the effects of NED \cite{stuchlik2019shadow}. 

For the wave (null 4-momentum) vector $k_{\alpha}$, the geodesics equation $k_{\mu;\alpha}k^{\alpha}=0$ is complemented by the normalization condition $k^{\alpha}k_{\alpha}=0$. In the spherically symmetric spacetimes, the motion of test particles is planar, being realized in the central planes of the spacetimes. When the motion of a single particle is considered, the central plane can be conveniently chosen to be the equatorial plane of the spacetime coordinate system. If we have to explore numerous particles simultaneously, as is the case of the optical phenomena, we have to write the geodesic equation in a general central plane. Then the Hamiltonian method can be conveniently applied to treat null geodesics (see e.g. \cite{stuchlik2019shadow,stuchlik2018light}). 

\subsection{Neutrinos and photons null geodesics of the spacetime geometry}

The neutrinos are considered to be massless (null) particles that are not affected by the NED effects and follow null geodesics of the spacetime geometry ~(\ref{eq:geometry}). Consequently, they serve as the reference for analyzing the optical phenomena connected with the propagation of photons, influenced by the non-linearities of electrodynamics that can be reflected in both the trajectory and the frequency shift. 

There are two conserved constants of the geodesic motion associated with the cyclic coordinates $\phi$, $t$ 
\begin{equation}
	L_z=k_\phi\ ,\quad\textrm{and}\quad E=-k_t\ .
	\label{consts}
\end{equation}

Moreover, there is a constant of motion associated with the separation constant arising during the solution of the Hamilton-Jacobi equation that is denoted as $Q$ (see e.g. \cite{stuchlik2019shadow}). Further, we define impact parameters
\begin{equation}
	b\equiv\frac{L_z}{E},\quad\textrm{and}\quad q\equiv\frac{Q}{E^2}.
	\label{consts1}
\end{equation}  

Then the integrated null geodesics of the spacetime geometry (\ref{eq:geometry}) that will be used here for comparison with the motion of photons can be written in the form \cite{stuchlik2019shadow} 
\begin{eqnarray}
    k^r&=&\pm\sqrt{1-f(r)(b^2+q)/r^2},\\
    k^\theta &=& \pm\frac{1}{r^2}\sqrt{q - b^2 \cot^2 \theta},\\
    k^\phi &=& \frac{b}{r^2\sin^2\theta},\\
    k^t &=&\frac{1}{f(r)}.
\end{eqnarray}

These equations are expressed for the motion of neutrino in a general central plane. Here we study the motion in the equatorial plane $\theta=const=\pi/2$. The quantity $I^2 = L_{z}^2 + Q$ represents the total angular momentum of the particle, and $\mathcal{I}^2 = b^2 + q$ represents the total impact parameter. 

The motion of photons in Schwarzschild or Reissner-Nordstrom spacetime is also governed by null geodesics of the spacetime, and the equation of motion can be obtained in the same way how it was done for the equation of motion for neutrinos presented in this subsection, with the appropriate choice of the lapse function ~(\ref{eq:lapse_1}) or (\ref{eq:lapse_2}) in each case.

For further calculations, it is convenient to introduce a new radial coordinate $u \equiv 1/r$. Then the line element reads
\begin{equation}
	\diff s^2 = -\tilde{f}(u) \diff t^2 + \frac{1}{u^4 \tilde{f}(u)} \diff u^2 + \frac{1}{u^2} \diff \theta^2 + \frac{\sin^2 \theta}{u^2} \diff\phi^2\ ,	
	\label{eq:geometry_0}
\end{equation}
where function $\tilde{f}(u)$ can be obtained by substitution $r \rightarrow 1/u$ in the expression for the lapse functions  ~(\ref{eq:lapse_1}), ~(\ref{eq:lapse_2}) and ~(\ref{eq:lapse_3}):  
\begin{equation}
	\tilde{f}(u) = \left\{\begin{array}{cc}
	1- 2 \, M \, u & \textrm{ for } Schw \ ,\\
	1- 2 \, M \, u + q_m^2 \, u^2 & \textrm{ for } RN \ ,\\
	1-\frac{2 \, M \, u}{(1 + q_m \, u)^3} & \textrm{ for }GRBHNED\ .
	\end{array}\right.
	\label{eq:lapse_u}
\end{equation}

In the last expression we have introduced notion $Schw$ for Schwarzschild black hole, $RN$ for Reissner-Nordstrom black hole and $GRBHNED$ for generic regular black hole related to non-linear electrodynamics with Maxwelian weak-field limit. The integrated null geodesic equations of the spacetime geometry then read  
\begin{eqnarray}
    k^u&=&\pm u^2\sqrt{1-\tilde{f}(u)(b^2+q)u^2},\\
    k^\theta &=& \pm u^2\sqrt{q - b^2 \cot^2 \theta},\\
    k^\phi &=& \frac{b \, u^2}{\sin^2\theta},\\
    k^t &=&\frac{1}{\tilde{f}(u)}.
\end{eqnarray}

\subsection{Photons null geodesics of the effective geometry}

Photons propagate along the null geodesic of the effective geometry that is spherically symmetric as the spacetime geometry. The detailed derivation of the effective geometry metric is provided in \cite{stuchlik2019shadow}. Here we just briefly remind the results which were obtained there.  

In \cite{stuchlik2019shadow} it was shown that the line element of effective geometry transforms as follows  

\begin{equation}
	\diff s_{eff}^2=-\frac{f(r)}{\mathcal{L}_F}\diff t^2+\frac{1}{\mathcal{L}_F f(r)}\diff r^2 + \frac{r^2}{\Phi}(\diff\theta^2+\sin^2\theta\phi^2)\ , 
	\label{eq:max_geometry}
\end{equation}
where 
\begin{equation}
\mathcal{L}_F\equiv\frac{\diff \mathcal{L}}{\diff F}\ ,\quad\mathcal{L}_{FF}\equiv\frac{\diff^2\mathcal{L}}{\diff F^2} \ , 
\end{equation}
and
\begin{equation}
    \Phi\equiv\mathcal{L}_F+2\mathcal{L}_{FF}F.
\end{equation}

The photon motion is again restricted to the central planes and can be treated by the Hamilton method. There are conserved constants of the motion associated with the cyclic coordinates $t$ and $\phi$ of the effective geometry that enter the corresponding Hamiltonian \cite{stuchlik2019shadow} 
\begin{equation}
	H=\frac{1}{2}g^{\mu\nu}_{eff}k_\mu k_\nu.
\end{equation}

The corresponding constants of motion and impact parameters are defined in analogy to formulae ~(\ref{consts}) and ~(\ref{consts1}). The integrated geodesic equations for the photon motion in the effective geometry can be then written in the form (see also \cite{stuchlik2019shadow}): 
\begin{eqnarray}
	\left[k^r\right]^2 &=&\mathcal{L}_F\left(\mathcal{L}_F-\frac{\Phi f}{r^2}(b^2+q)\right), \label{radial}\\
	\left[k^\theta\right]^2 &=& \frac{\Phi^2}{r^4}\left(q - b^2 \cot^2 \theta \right),\\
	k^t &=& \frac{\mathcal{L}_F}{f(r)},\\
	k^\phi &=& \frac{\Phi\,b}{r^2\sin^2\theta}.
	\label{eq:int_geo_eq}
\end{eqnarray}

These equations are written in a general central plane -- as in the case of the motion in the spacetime geometry, we have the total impact parameter of the photon determined as $\mathcal{I}^2 = b^2 + q$. 

For further calculations, it is convenient to introduce a new radial coordinate $u \equiv 1/r$. With such a replacement, the line element in spacetime related to NED reads
\begin{equation}
	\diff s_{eff}^2 = -\frac{\tilde{f}(u)}{\mathcal{L}_F} \diff t^2 + \frac{1}{u^4 \tilde{f}(u)\mathcal{L}_F} \diff u^2 + \frac{1}{u^2 \Phi} \diff \theta^2 + \frac{\sin^2 \theta}{u^2 \Phi} \diff\phi^2.	
	\label{eq:geometry_1}
\end{equation}

The integrated null geodesic equations of the effective geometry then read 
\begin{eqnarray}
      \tilde{k}^{u} &=& \pm u^2 \mathcal{L}_F \sqrt{(1 - \frac{\Phi}{\mathcal{L}_F} \tilde{f}(u) (b^2+q)u^2)},\\ 
      \tilde{k}^{\theta} &=& \pm u^2 \Phi \sqrt{q - b^2 \cot^2 \theta},\\
      \tilde{k}^{t} &=&  \tilde{g}^{tt} \tilde{k}_t  = \frac{\mathcal{L}_F}{\tilde{f}(u)},\\
      \tilde{k}^{\phi} &=&  \tilde{g}^{\phi\phi} \tilde{k}_{\phi}  = \frac{\Phi u^2}{\sin^2 \theta}b\ ,
      \label{eq:int_geo_eq_1}
\end{eqnarray}
where contravariant components of the effective geometry metric tensor are given by 
\begin{equation}
    \tilde{g}^{\alpha\beta} = \mathcal{L}_F g^{\alpha\beta} - \mathcal{L}_{FF} F^{\alpha}_{\gamma} F^{\gamma\beta} \ . 
\end{equation}

\section{\label{sec:light-curves}Construction of light curves from hot spots}

To study the effects of a strong gravitational field in the vicinity of a black hole related to NED, we explore radiation from a small hot spot orbiting  in a Keplerian circular orbit. For radiation propagating in a strong gravitational field, it is necessary to take into account the relativistic effects that affect this radiation. Consequently, to obtain the light curve, it is necessary to study the focusing of the radiation (gravitational lensing), the frequency shift of the radiation, and the time delay.

\subsection{\label{sec:grav-lens}Gravitational lensing}

Magnification of the radiation flux can be described by a solid angle at which a distant observer can see a hot spot. Such solid angle is defined in coordinates related to the image of the hot spot on the observer's detector as follows
\begin{equation}
    \diff \Omega = \frac{1}{d_o^2} \, b \, \diff b \, \diff \phi^{\prime}\ ,
    \label{eq:solid_angle_1}
\end{equation}
where $d_o$ is the distance from the hot spot to the observer, $b$ is the impact parameter, $\phi^{\prime}$ is the angle defined on the image of the hot spot on the observer's detector. 

The impact parameter and the angle on the hot spot image are related to the radial $u_s = 1/r_s$ and azimuthal $\phi_s$ coordinates that determine the position of the hot spot in the considered spacetime. Using the Jacobian of transformation between these two coordinate systems 
\begin{equation}
    J = \left| \frac{\partial b}{\partial u_s}\frac{\partial \phi^{\prime}}{\partial \phi_s} - \frac{\partial b}{\partial \phi_s} \frac{\partial \phi^{\prime}}{\partial u_s} \right|,
    \label{eq:jacob}
\end{equation}
one can rewrite equation ~(\ref{eq:solid_angle_1}) as follows
\begin{equation}
    \diff \Omega = \frac{1}{d_o^2} \, J \, b \, \diff u_s \, \diff \phi_s.
    \label{eq:solid_angle_2}
\end{equation}

From the spherical geometry, one can get the relationship between the angles $\phi_s$ and $\phi^{\prime}$ as 
\begin{equation}
    \tan{\phi^{\prime}} = \frac{\tan{\phi_s}}{\cos{\theta_o}} \ , 
    \label{eq:rel_phi_phi}
\end{equation}
where $\theta_o$ is the latitudinal angle that determines the observer's position in the considered spacetime. Applying the implicit function theorem to the last expression, one can find the expression for the partial derivative $\partial \phi^{\prime} / \partial \phi_s$ as follows 
\begin{equation}
    \frac{\partial \phi^{\prime}}{\partial \phi_s} = \frac{\cos{\theta_o}}{1-\cos^2{\phi_s} \, \sin^2{\theta_o}}\ .
    \label{eq:par_dif_phi_phi}
\end{equation}
Since the angle $\phi^{\prime}$ depends implicitly only on $\phi_s$, and does not depend on $b$, then the partial derivative $\partial \phi^{\prime}/\partial b = 0$, and the Jacobian of the transformation ~(\ref{eq:jacob}), is reduced to

\begin{equation}
    J = \left| \frac{\partial b}{\partial u_s}\frac{\partial \phi^{\prime}}{\partial \phi_s} \right|.
    \label{eq:jacob_red}
\end{equation}

In order to find the partial derivative $\partial b / \partial u_s$, we use the integrated geodesic equations, which describe photon motion in the radial and azimuthal directions ~(\ref{eq:int_geo_eq_1}). From these two equations, dividing one by the other and integrating from $\phi_s$ to $\phi_o$ and from $u_s$ to $u_o$, one can find the dependence of the angle $\varphi$, which sweeps out the photon propagating in the central plane from the hot spot to the observer, on the radial coordinate of the hot spot and observer. It is worth paying attention to the fact that depending on the photon's trajectory, turning points in the radial coordinate must be taken into account when integrating from $u_s$ to $u_o$. Accordingly, in the absence or in the presence of one turning point, the expression for the angle $\varphi$ takes the form

\begin{equation}
	\varphi=\left\{\begin{array}{cc}
	\int_{u_{o}}^{u_{s}}\frac{\diff u}{\sqrt{U(u, b)}} & \textrm{ for }n_u=0,\\
	\int_{u_{o}}^{u_{t}(b)}\frac{\diff u}{\sqrt{U(u, b)}}+\int_{u_{s}}^{u_{t}(b)}\frac{\diff u}{\sqrt{U(u, b)}} & \textrm{ for } n_u=1.
	\end{array}\right.
	\label{eq:phi_center_plane}
\end{equation}

For convenience in further calculations, we introduced the notation $U(u, b)$ for the function under the integral in the previous expression. This function for different spacetimes takes the following forms
\begin{equation}
	U(u, b) \equiv \left\{\begin{array}{cc}
	2 u^3 - u^2 +\frac{1}{b^2} & \textrm{ for }Schw\ ,\\
	2 u^3 - u^2 - q_m^2 u^4 +\frac{1}{b^2} & \textrm{ for } RN\ ,\\
	\frac{\mathcal{L}_F^2}{b^2 \Phi^2}-\frac{u^2 \, \tilde{f}(u)}{\Phi \, \mathcal{L}_F} & \textrm{ for }GRBHNED\ .
	\end{array}\right.
	\label{eq:func_u_big}
\end{equation}

In case when there is no turning point in the radial direction ($n_u=0$), the photon moving along such geodesic has the impact parameter $b$ that is less than impact parameter $b_{ph}$ being appropriate to the photon on photon's circular orbit. On the other hand, in the case of one turning point in the radial direction ($n_u=1$), the impact parameter $b$ for the photon following such trajectory is higher than the impact parameter $b_{ph}$ for the photon on the circular photon orbit.        

Next, we will choose the orientation of the coordinate axes in such a way that the position of the hot spot is determined by the coordinates ($u_s, \theta_s = \pi/2, \phi_s $) and the position of the observer ($u_o, \theta_o, \phi_o = 0 $). From the spherical geometry, one can write the relation between $\phi_s$ and $\varphi$ as 
\begin{equation}
    \cos{\varphi} = \sin{\theta_o} \, \cos{\phi_s}.
\end{equation}

Thus, the angle $\varphi$ swept by the photon when moving in the central plane from the hot spot to the observer is constant for the considered azimuthal positions of the source $\phi_s$ and latitudinal position of the observer $\theta_o$. Taking advantage of this fact and applying it in the equation ~(\ref{eq:phi_center_plane}), one can introduce a function that implicitly determines the relationship between $u_s$ and $b$ as follows 
\begin{equation}
    F_1(u_s, b) = \varphi(\theta_o, \phi_s) - \int_{u_{o}}^{u_{s}}\frac{\diff u}{\sqrt{U(u, b)}} = 0,
    \label{eq:f1}
\end{equation}
in case $b < b_{ph}$, and 

\begin{equation}
    F_2(u_s, b) = \varphi(\theta_o, \phi_s) - \int_{u_{o}}^{u_{t}(b)}\frac{\diff u}{\sqrt{U(u, b)}}-\int_{u_{s}}^{u_{t}(b)}\frac{\diff u}{\sqrt{U(u, b)}} = 0,
    \label{eq:f2}
\end{equation}
in case $b > b_{ph}$. 

Applying the theorem on the derivative of a function given implicitly to the introduced functions, one can obtain the required expression for the partial derivative $\partial u_s / \partial b$ in two cases as follows 
\begin{equation}
	\frac{\partial u_s}{\partial b}=\left\{\begin{array}{cc}
	-\frac{\partial F_1 / \partial b}{\partial F_1 / \partial u_s} & \textrm{ for } b < b_{ph},\\
	-\frac{\partial F_2 / \partial b}{\partial F_2 / \partial u_s} & \textrm{ for } b > b_{ph}.
	\end{array}\right.
\end{equation}
Thus for photons moving along null geodesics with impact parameter $b < b_{ph}$
\begin{equation}
    \frac{\partial u_s}{\partial b}\bigg\vert_{b < b_{ph}} = \sqrt{U(u_s, b)} \int_{u_{o}}^{u_{s}}\frac{\partial U / \partial b}{2 U^{3/2}}\diff u,
    \label{eq:dus_db_1}
\end{equation}
and for photons moving along null geodesics with impact parameter $b > b_{ph}$
\begin{equation}
    \begin{split}
    \frac{\partial u_s}{\partial b}\bigg\vert_{b > b_{ph}} &= - \sqrt{U(u_s, b)} \left[ \int_{u_{o}}^{u_{t}(b)}\frac{\frac{\partial P}{\partial b} + \frac{\partial u_t}{\partial b} \frac{\partial P}{\partial u}}{2 P \sqrt{U}}\diff u + \int_{u_{s}}^{u_{t}(b)}\frac{\frac{\partial P}{\partial b} + \frac{\partial u_t}{\partial b} \frac{\partial P}{\partial u}}{2 P \sqrt{U}}\diff u \right. \\
    &\left. - \frac{\partial u_t}{\partial b} \left( \frac{1}{\sqrt{U(u_o, b)}} + \frac{1}{\sqrt{U(u_s, b)}} \right) \right].
    \label{eq:dus_db_2}
    \end{split}
\end{equation}
In the last expression, we introduced the new function
\begin{equation}
    P(u, b) \equiv \frac{U(u, b)}{u_t - u}.
\end{equation}

The introduction of such function makes it possible to solve the problems with the divergence of the integrals at the turning point, which may arise in case of direct differentiation of the function $F_2(u_s, b)$ in ~(\ref{eq:f2}) with respect to the impact parameter $b$. Detailed derivation of the expressions ~(\ref{eq:dus_db_1}) and ~(\ref{eq:dus_db_2}) is presented in the Appendix \ref{sec:app-1}. The partial derivatives of the function $P(u, b)$ in case of $u \neq u_t$ are 
\begin{eqnarray}
    \frac{\partial P}{\partial u} &=& \frac{1}{u_t - u} \left( \frac{\partial U}{\partial u} + \frac{U}{u_t - u}\right),\\
	\frac{\partial P}{\partial b} &=& \frac{1}{u_t - u} \left( \frac{\partial U}{\partial b} + \frac{U}{u_t - u} \frac{\partial u_t}{\partial b}\right).
\end{eqnarray}
In case of $u = u_t$ the partial derivatives of the function $P(u, b)$ are 
\begin{eqnarray}
    \lim_{u \to u_t} \frac{\partial P}{\partial u} &=& -\frac{1}{2} \frac{\partial^2 U}{\partial u^2}\bigg\vert_{u_t},\\
	\lim_{u \to u_t} \frac{\partial P}{\partial b} &=& \frac{\partial P}{\partial u}\bigg\vert_{u_t} \frac{\partial u_t}{\partial b}.
\end{eqnarray}

The turning point $u_t$ is determined by the value of the impact parameter $b$. Note that at the turning point, the function $U$ is equal to zero
\begin{equation}
    U(u_t, b) = 0.
\end{equation}
Thus applying the theorem on the derivative of an implicitly defined function in the last expression, we obtain the partial derivative $\partial u_t / \partial b$ as follow
\begin{equation}
    \frac{\partial u_t}{\partial b} = -\frac{\partial U/ \partial b}{\partial U/ \partial u_t}.
\end{equation}

When we have find the partial derivatives $\partial b / \partial u_s$ and $\partial \phi^{\prime} / \partial \phi_s$ which are necessary to calculate the Jacobian of transformation ~(\ref{eq:jacob_red}), we have everything to determine the solid angle ~(\ref{eq:solid_angle_2}) in which the distant observer sees the hot spot.

\subsection{\label{sec:freq-shift}Frequency shift}

In order to obtain the radiation flux ~(\ref{eq:flux}) detected by a distant observer, it is necessary to find the frequency shift that occurs due to the influence of a strong gravitational field, as well as due to the Doppler effect arising from the movement of the hot spot.

Since the photon energy $E=h\nu$ is related to the radiation frequency $\nu$, expression for the frequency shift ~(\ref{eq:red-factor}) can be rewritten as follow
\begin{equation}
    g=\frac{E_o}{E_s}=\frac{\left(-k_{\mu} U^{\mu}\right)_o}{\left(-k_{\mu} U^{\mu}\right)_s},
\end{equation}
where $E_s$ is the energy of the photon in place of emission, $E_o$ is the energy of the photon in place of detection,   $k_{\mu}$ is the wave vector, $(U^{\mu})_s=(U^t,0,0,U^{\phi})_s$ and $(U^{\mu})_o=(U_t,0,0,0)_o$ are the four velocities of the hot spot and observer. Further, using the normalization condition for the four-velocity vector $U_{\mu}U^{\mu}=-1$, and assuming that the observer is at a sufficient distance from the black hole where space is flat, one can obtain the expression for the frequency shift in case of the Schwarzschild, Reissner-Nordstrom and Maxwellian regular black holes 
\begin{equation}
	g = \left\{\begin{array}{cc}
	\frac{\sqrt{f_{Sch}(u_s) - \frac{\omega_K^2(u_s)}{u_s^2}}}{1 -  l(b) \, \omega_K^2(u_s)} & \textrm{ for }Schw \ ,\\
	\frac{\sqrt{f_{RN}(u_s) - \frac{\omega_K^2(u_s)}{u_s^2}}}{1 -  l(b) \, \omega_K^2(u_s)} & \textrm{ for } RN\ ,\\
	\frac{\mathcal{L}_F(u_o)}{\mathcal{L}_F(u_s)} \frac{\sqrt{\tilde{f}(u) - \frac{\omega_K^2(u_s)}{u_s^2}}}{1 - \frac{\Phi(u_s)}{\mathcal{L}_F(u_s)} \, l(b) \, \omega_K^2(u_s)} & \textrm{ for }GRBHNED\ , 
	\end{array}\right.
	\label{eq:freq_shift}
\end{equation}
where $\omega_K(u)$ is the angular velocity of the hot spot moving in the Keplerian circular orbit defined by the general formula $\omega_K(u) \equiv U^\phi/U^t$. The expression for the angular velocity in static and spherically symmetric spacetime reads
\begin{equation}
    \omega^2_K(u)=\frac{g_{tt,u}}{g_{\phi\phi,u}}=\frac{u^3}{2}f(u)_{,u}.
\end{equation}
After substituting appropriate components of the metric tensor of Schwarzschild, Reissner-Nordstrom, or Maxwellian regular black hole, expressions for the angular velocity take form  
\begin{equation}
	\omega_K(u) = \left\{\begin{array}{cc}
	\left( M u^3\right)^{1/2} & \textrm{ for }Schw\ ,\\
	\left( M u^3 - q_m^2 u^4\right)^{1/2} & \textrm{ for } RN\ ,\\
	\left[ \frac{M\left(\frac{1}{u} - 2q_m\right)}{\left(\frac{1}{u} + q_m\right)^4}\right]^{1/2} & \textrm{ for }GRBHNED\ .
	\end{array}\right.
	\label{eq:ang_vel_3}
\end{equation}

\subsection{\label{sec:time-delay}Time delay for photons}

For each individual photon, the moment in time when it was emitted is related to the angular position and angular velocity of the hot spot as follow
\begin{equation}
    t_{e(i)}=\frac{\phi_{s(i)}}{\omega_K(u_{s})}.
\end{equation}

Since different hot spot positions in a circular orbit correspond to different values of the impact parameter $b$, the time delay for photons emitted from different positions of the hot spot in a circular orbit also depends on the impact parameter $b$. Relation between time delay and impact parameter $b$, can be found using equations of motion for $t$ and $u$ coordinates, dividing one by the other and integrating from $t_e$ to $t_o$ and from $u_s$ to $u_o$. The time delay is determined by the integral
\begin{equation}
	\Delta t_{(i)}=\left\{\begin{array}{cc}
	\int_{u_{o}}^{u_{s}} H(u, b)\diff u & \textrm{ for }n_u=0,\\
	\int_{u_{o}}^{u_{t}(b)} H(u, b)\diff u+\int_{u_{s}}^{u_{t}(b)} H(u, b)\diff u & \textrm{ for } n_u=1,
	\end{array}\right.
	\label{eq:time_delay}
\end{equation}
where $n_u$ is the number of the turning points in the radial direction for the photon moving along geodesic with impact parameter $b$. The function $H(u, b)$ takes for the considered metrics the following form
\begin{equation}
	H(u, b) \equiv \left\{\begin{array}{cc}
	\frac{1}{u^2 \left(1 - 2u\right) \sqrt{1 - b^2 u^2 \left(1 - 2u\right)}} & \textrm{ for }Schw \ ,\\
	\frac{1}{u^2 \left(1 - 2u + q_m^2 u^2\right) \sqrt{1 - b^2 u^2 \left(1 - 2u + q_m^2 u^2\right)}} & \textrm{ for } RN\ ,\\
	\frac{1}{u^2 \left[1-\frac{2 u}{(1 + q_m u)^3}\right] \sqrt{1-b^2 u^2 \frac{\Phi}{\mathcal{L}_F} \left[1-\frac{2 u}{(1 + q_m u)^3} \right]}} & \textrm{ for }GRBHNED \ .
	\end{array}\right.
	\label{eq:func_h_big}
\end{equation}

As a result, the observation time is the sum of the radiation time $t_{e(i)}$ and the time delay $\Delta t_{(i)}$ for each photon moving from the source to the observer

\begin{equation}
    t_{o(i)}=t_{e(i)} + \Delta t_{(i)}\ .
\end{equation}

\section{\label{sec:results}Main results}

We suppose that the hot spots are moving on circular orbits around a regular  black hole. The distant observer is static and has fixed azimuthal and radial coordinates $\phi_o = 0$ and $u_o = 1/r_o = 0.0001$. The radius of the hot spot orbits can take different values in different spacetimes and for the different values of the black hole parameters. We have assumed the hot spot orbits defined independently on coordinate systems, or those having physical properties that can be compared in different spacetimes. We have calculated the frequency shift, gravitational lensing and have constructed the light curves of the hot spots on ISCOs and of the hot spots with equal orbital periods for different values of the charge parameter $q_m$ moving around Maxwellian regular black hole.  Also, we have compared them with the results obtained for the hot spots moving with equal orbital periods around Schwarzschild and Reissner-Nordstrom black holes.

\begin{figure}[H]
	\begin{center}
		\begin{tabular}{cc}
			\includegraphics[width=2.5in]{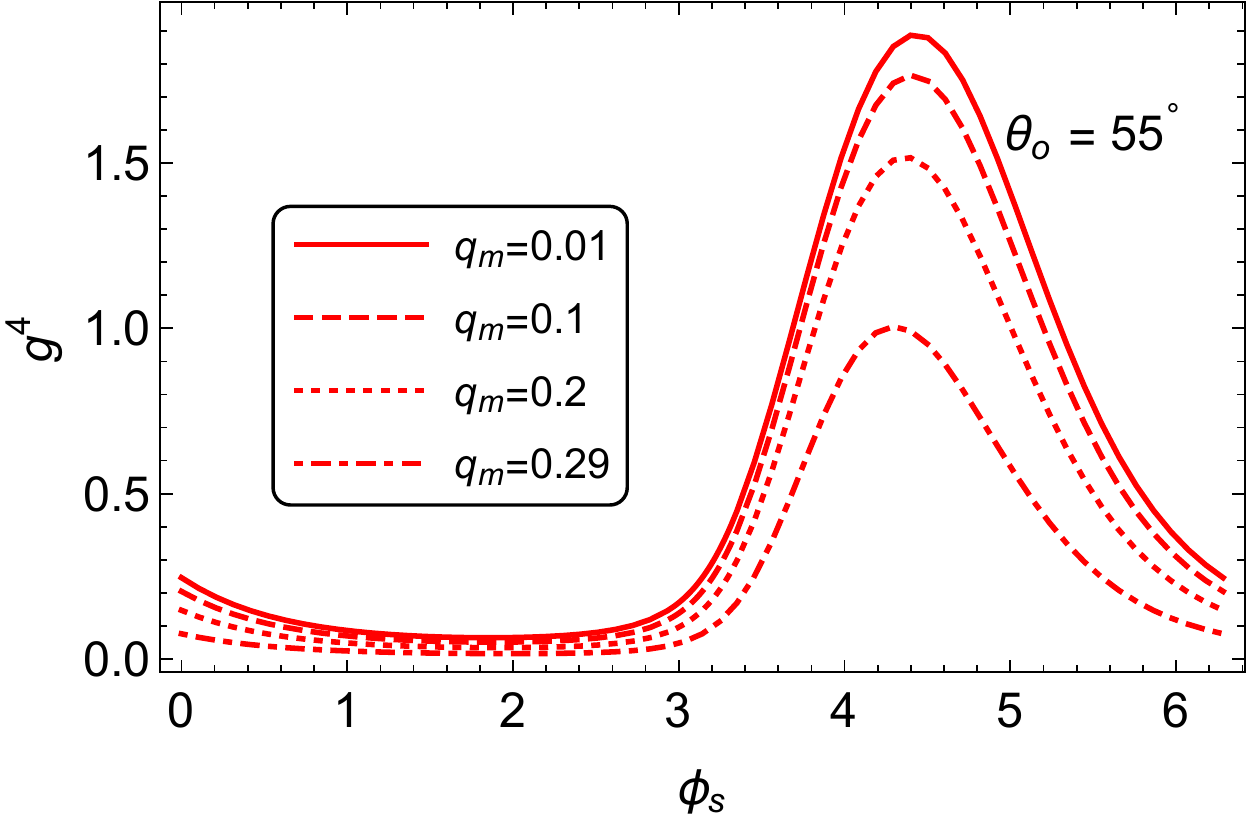}&\includegraphics[width=2.5in]{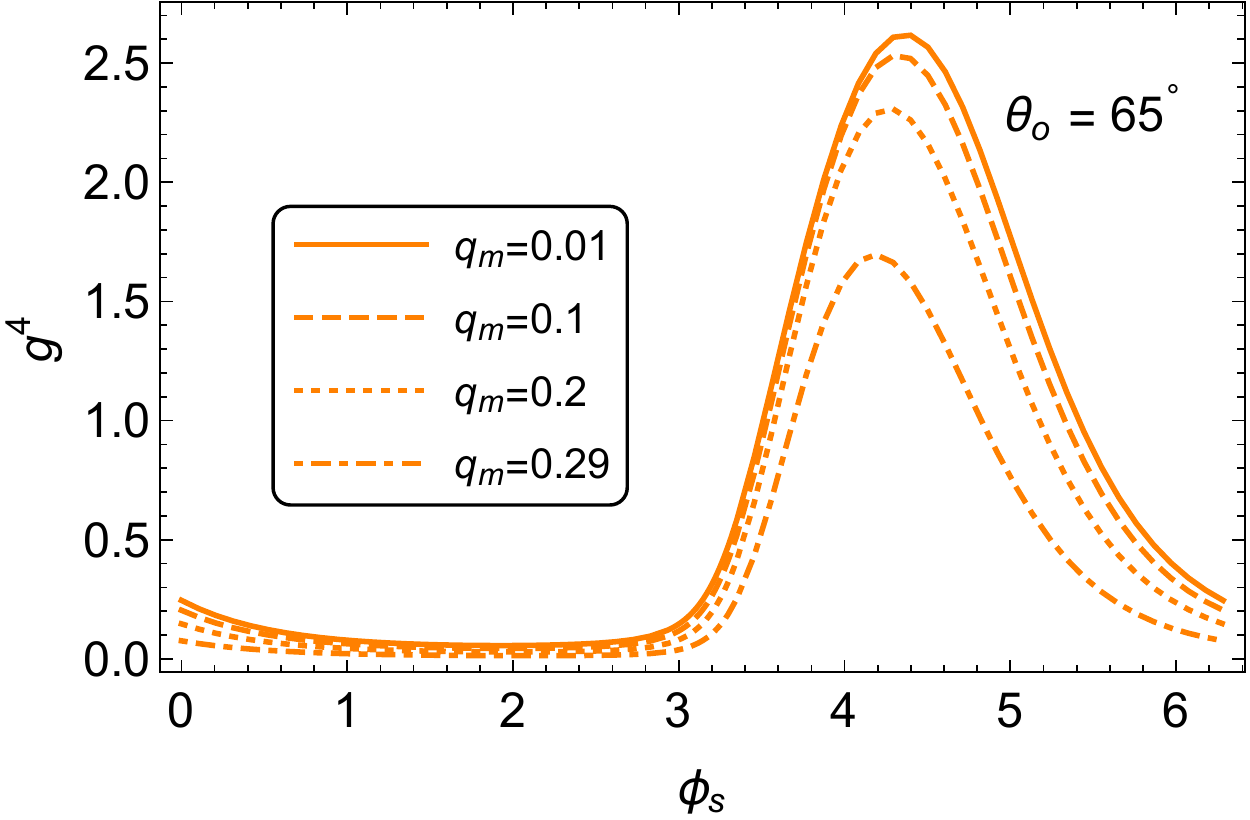}\\
			\includegraphics[width=2.5in]{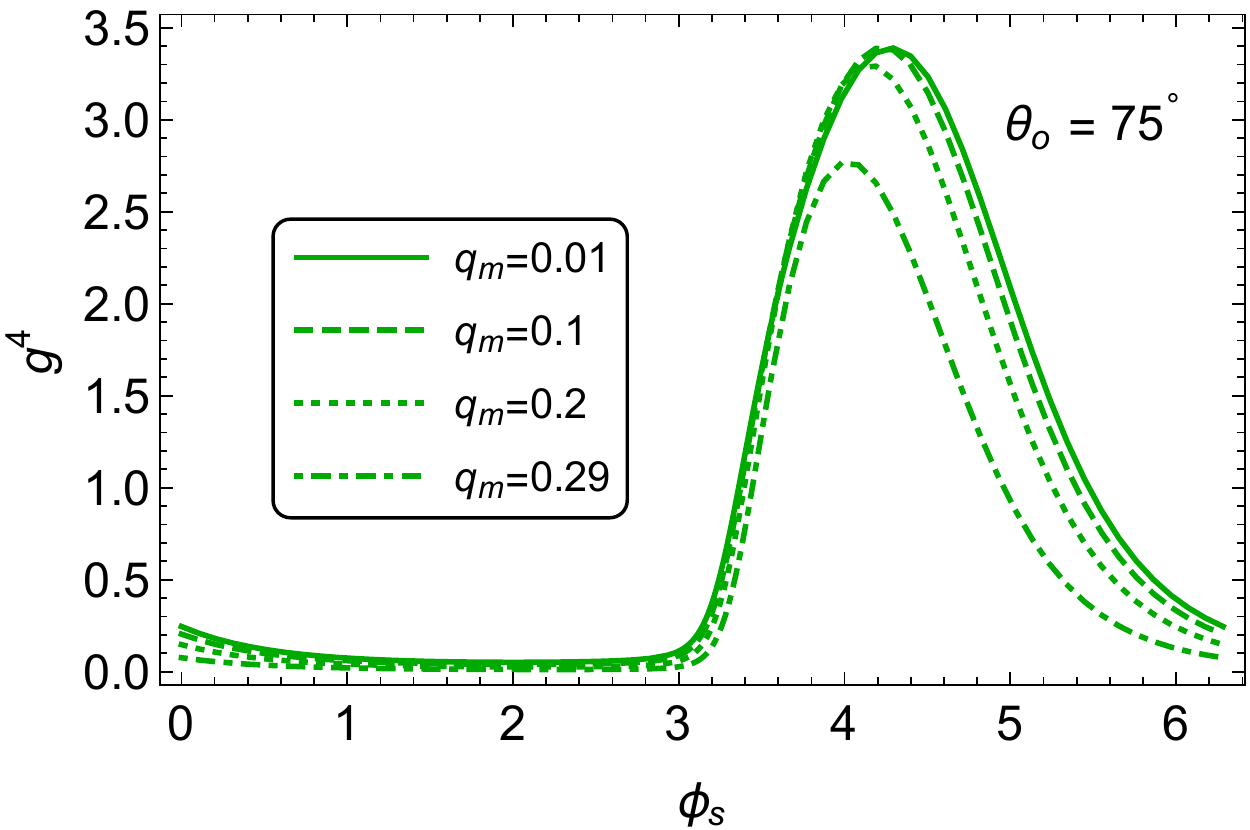}&\includegraphics[width=2.5in]{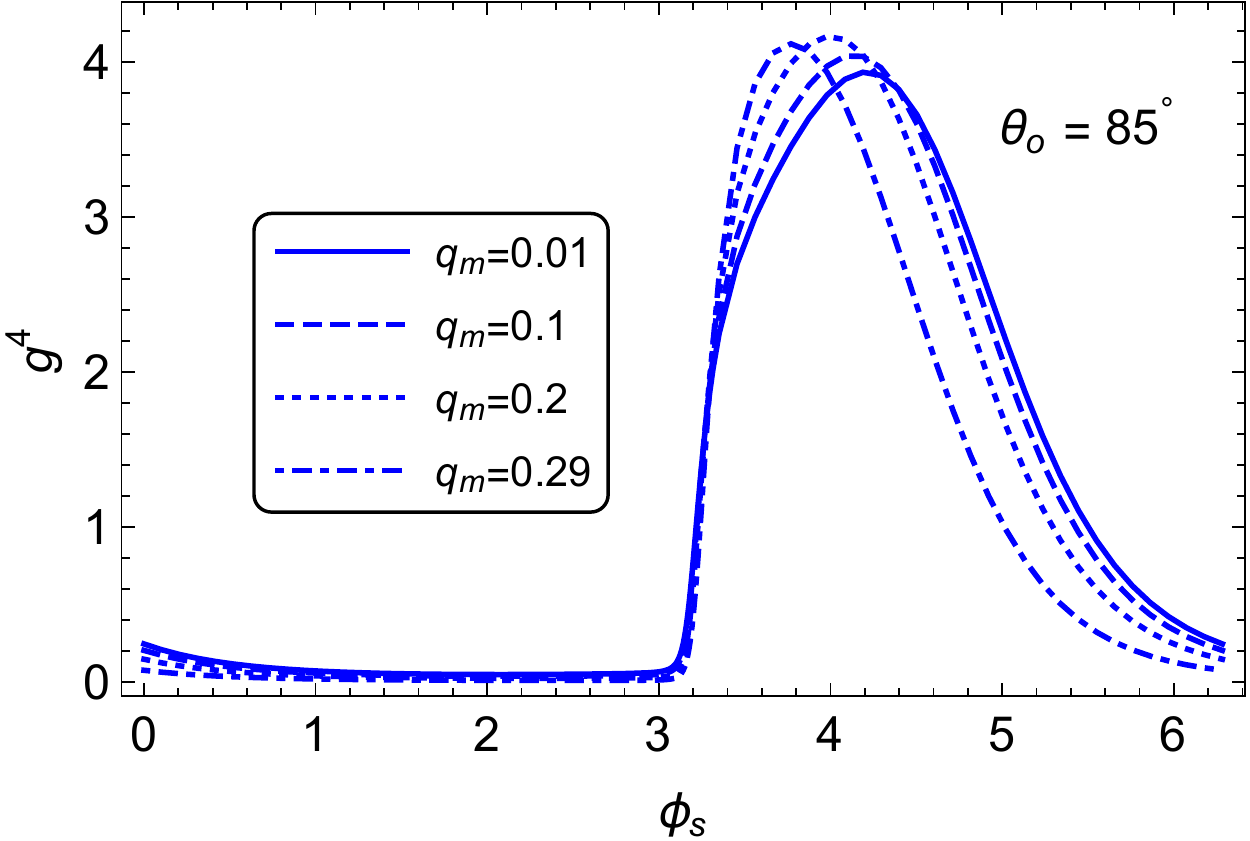}\\
		\end{tabular}
		\caption{The frequency shift $g^4$ as function of hot spots angular position $\phi_s$, for some different values of the inclination angle of the observer $\theta_o$. The spots are located on the ISCOs, which radii depend on the values of the parameter $q_m$. The units for frequency shift are arbitrary.}
		\label{fig:freq_shift}
	\end{center}
\end{figure}

The frequency of radiation detected by distant observers constantly changes due to effect of the two factors, gravitational redshift and the Doppler effect. The figure ~\ref{fig:freq_shift} shows the frequency  shift $g^4$ as  function  of  hot  spots  angular  position $\phi_s$ for four different inclination angles of the observer $\theta_o = 55^{\circ}, 65^{\circ}, 75^{\circ}, 75^{\circ}$ and for four representative values of the charge parameter $q_m = 0.01, 0.1, 0.2, 0.29$. The hot spots are located in ISCOs, however the radii of the ISCOs are different for different values of the charge parameter $q_m$. With increasing the inclination angle of the observer $\theta_o$, frequency shift $g^4$ also increases. With increasing the charge parameter $q_m$, the minimums of the frequency shift $q_m$ tend to bigger values of the hot spot angular position $\phi_s$ and maximum peaks tend to smaller values. The table ~\ref{tab:table1} presents the numerical values corresponding to the maxima and minima of the graphs.         

\newpage
\begin{table}[H]
\caption{\label{tab:table1}%
The columns present the inclination angle of the observer $\theta_o$, the charge parameter $q_m$, the radius of the hot spot circular orbit $r_s$, the hot spots angular positions $\phi_s$ corresponding to the maximum and minimum of the frequency shift $g^4$ and the maximum and minimum value of the frequency shift $g^4$. 
}
\begin{ruledtabular}
\begin{tabular}{ccccccc}
\textrm{$\theta_o$}&
\textrm{$q_m$}&
\textrm{$r_s$}&
\textrm{$\phi_s$}&
\textrm{$MAX(g^4)$}&
\textrm{$\phi_s$}&
\textrm{$MIN(g^4)$}\\
\colrule
 55 & 0.01 & 5.90968 & 4.43133 & 1.88864 & 1.85177 & 0.0640983 \\
 55 & 0.1 & 5.06204 & 4.40544 & 1.76569 & 1.8777 & 0.0506962 \\
 55 & 0.2 & 4.00497 & 4.36504 & 1.51803 & 1.91824 & 0.0340384 \\
 55 & 0.29 & 2.78485 & 4.30048 & 1.0053 & 1.98267 & 0.0160311 \\
\colrule
 65 & 0.01 & 5.90968 & 4.36019 & 2.62087 & 1.92306 & 0.0561272 \\
 65 & 0.1 & 5.06204 & 4.32437 & 2.53456 & 1.95872 & 0.0440818 \\
 65 & 0.2 & 4.00497 & 4.26868 & 2.3092 & 2.01463 & 0.0292618 \\
 65 & 0.29 & 2.78485 & 4.18146 & 1.69664 & 2.10185 & 0.0135162 \\
\colrule
 75 & 0.01 & 5.90968 & 4.28076 & 3.39046 & 2.00249 & 0.051078 \\
 75 & 0.1 & 5.06204 & 4.22931 & 3.39378 & 2.05382 & 0.0398489 \\
 75 & 0.2 & 4.00497 & 4.14765 & 3.29796 & 2.13559 & 0.0261404 \\
 75 & 0.29 & 2.78485 & 4.02186 & 2.772 & 2.26124 & 0.0118011 \\
\colrule
 85 & 0.01 & 5.90968 & 4.21076 & 3.93367 & 2.07234 & 0.0485488 \\
 85 & 0.1 & 5.06204 & 4.13566 & 4.04412 & 2.14753 & 0.0376761 \\
 85 & 0.2 & 4.00497 & 4.00268 & 4.16487 & 2.28044 & 0.0244445 \\
 85 & 0.29 & 2.78485 & 3.77254 & 4.11546 & 2.51086 & 0.0107319 \\
\end{tabular}
\end{ruledtabular}
\end{table}

\begin{figure}[H]
	\begin{center}
		\begin{tabular}{cc}
			\includegraphics[width=2.5in]{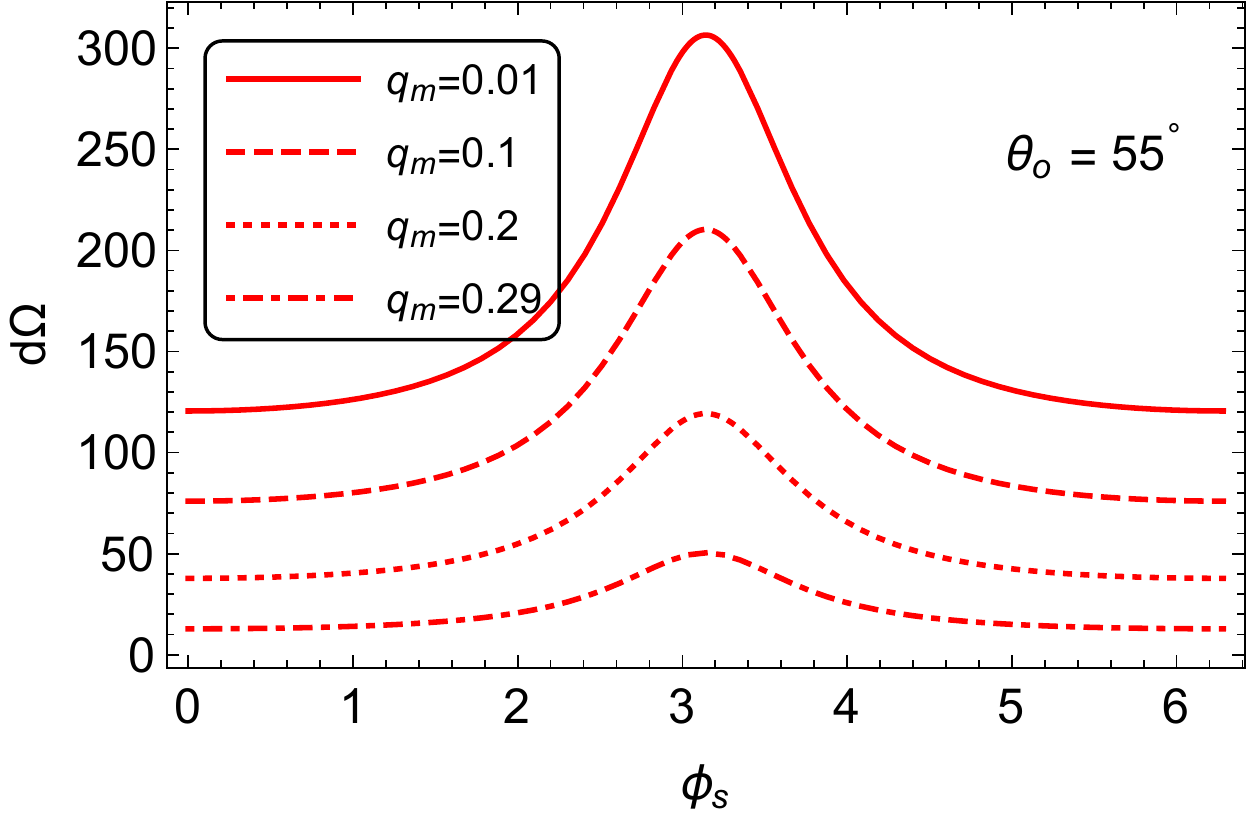}&\includegraphics[width=2.5in]{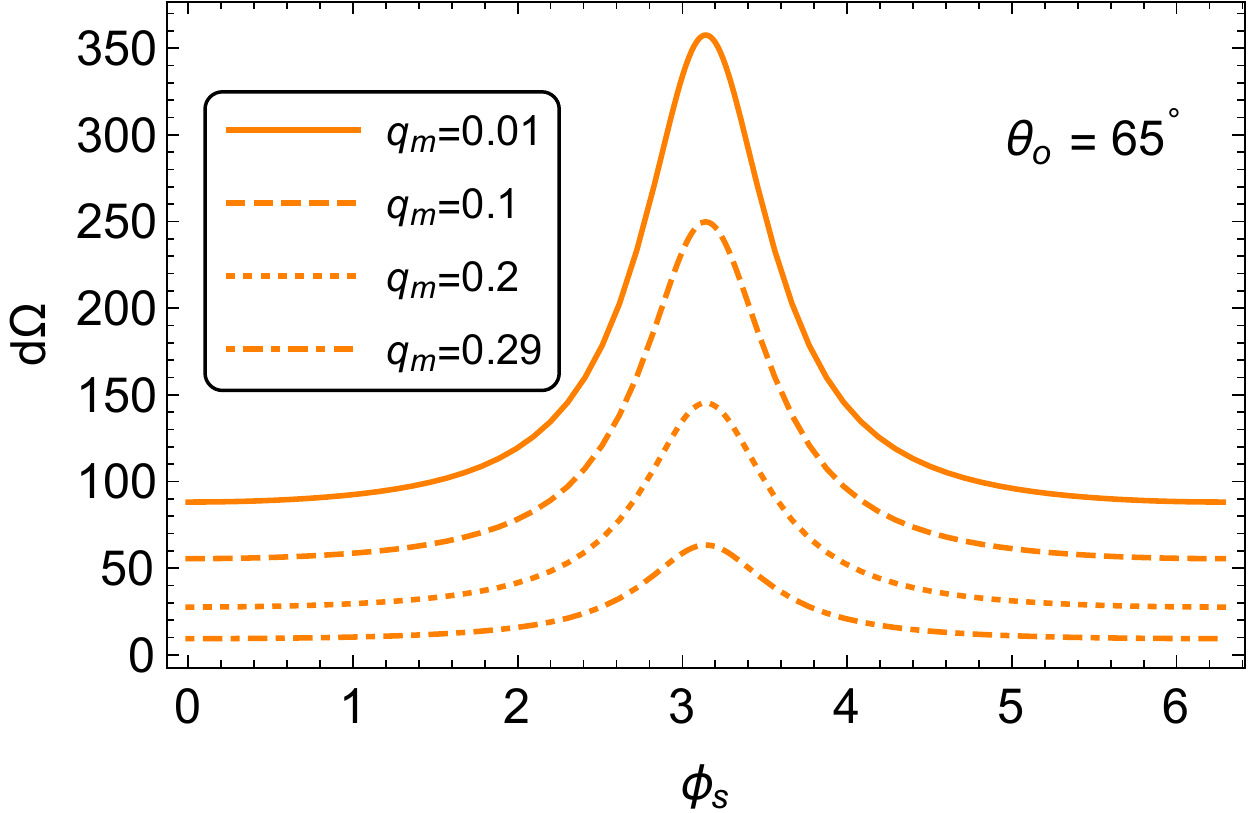}\\
			\includegraphics[width=2.5in]{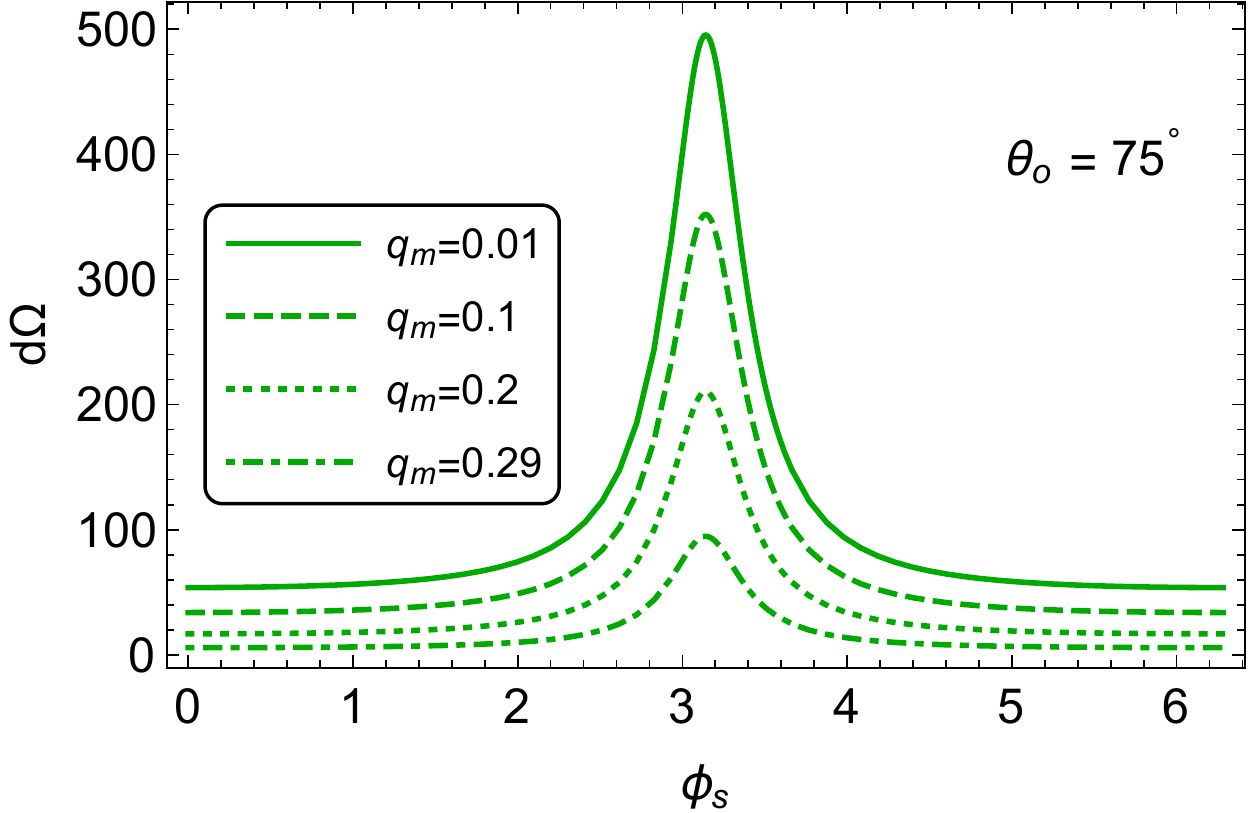}&\includegraphics[width=2.5in]{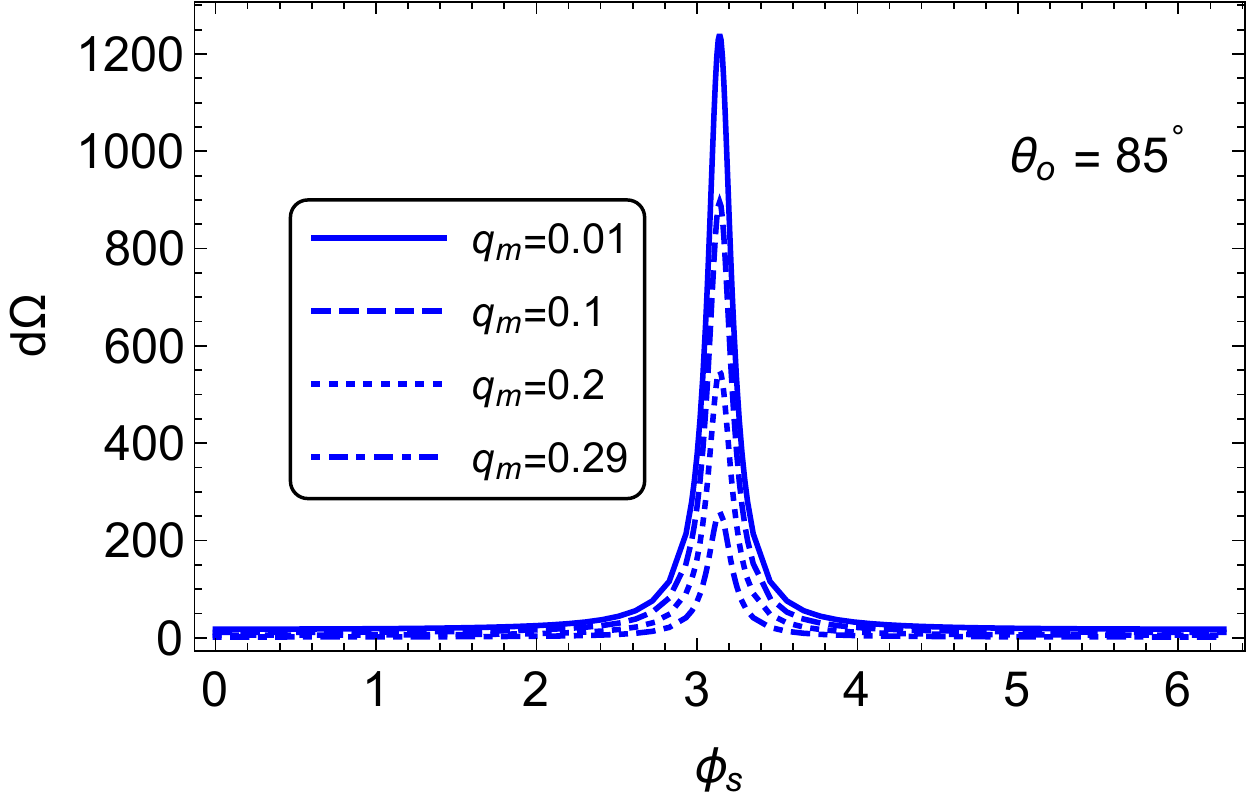}\\
		\end{tabular}
		\caption{The solid angle $\diff\Omega$ as function of hot spots angular position $\phi_s$, for some different values of the inclination angle of the observer $\theta_o$. The spots are located on the ISCOs, which radii depend on the values of the parameter $q_m$. The units for solid angle are arbitrary.}
		\label{fig:focusation}
	\end{center}
\end{figure}

\newpage
The effective area of the hot spot and the solid angle at which the distant observer would see the hot spot constantly changes as the hot spot moves around the black hole. The figure ~\ref{fig:focusation} demonstrates how the solid angle $\diff \Omega$ depends on the hot spots angular position $\phi_s$. For study we have selected four different inclination angles of the observer $\theta_o = 55^{\circ}, 65^{\circ}, 75^{\circ}, 75^{\circ}$ and for four representative values of the charge parameter $q_m = 0.01, 0.1, 0.2, 0.29$. The hot spots are located in ISCOs, that have different radii for different values of the parameter $q_m$. The maximal focusing occurs at $\phi_s = \pi$, when the hot spot is located behind the black hole. With increasing the inclination angle of the observer $\theta_o$, the effect of gravitational lensing becomes stronger.

\begin{figure}[H]
	\begin{center}
		\begin{tabular}{cc}
			\includegraphics[width=2.5in]{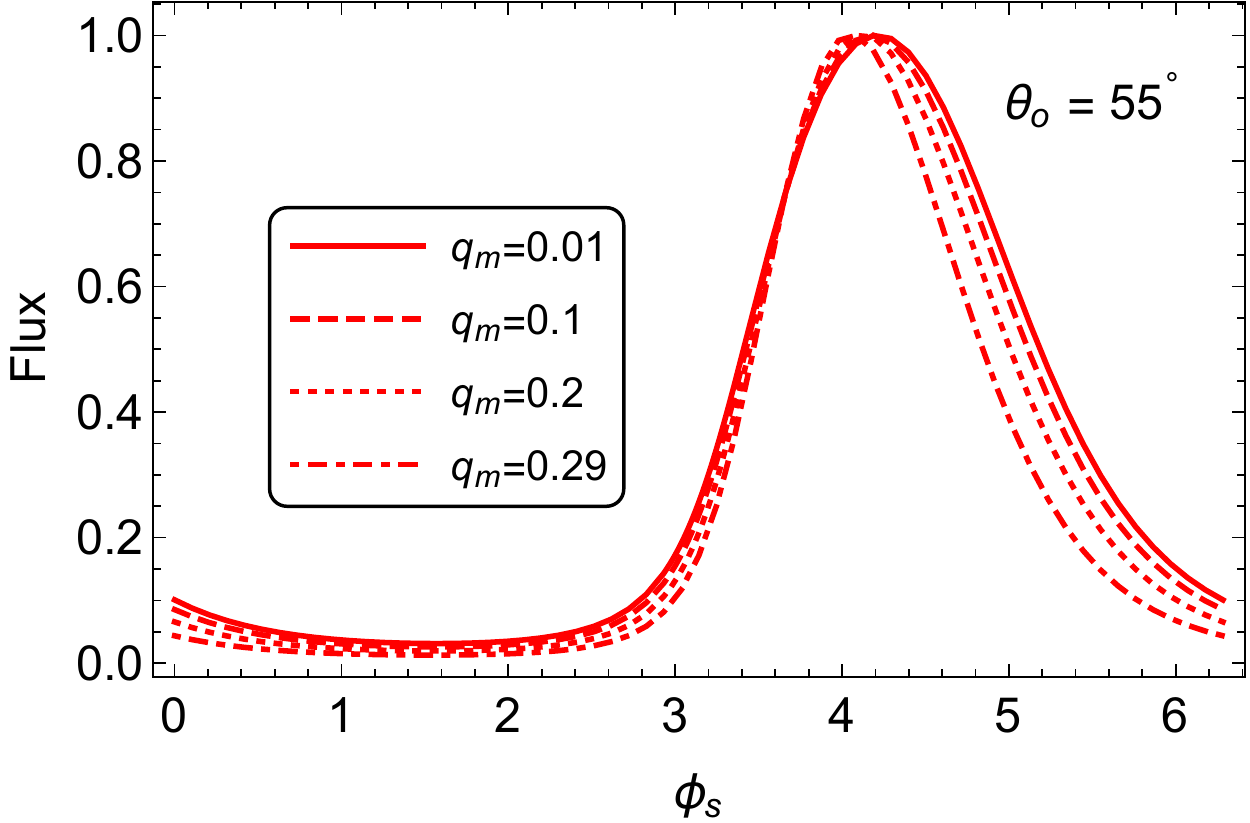}&\includegraphics[width=2.5in]{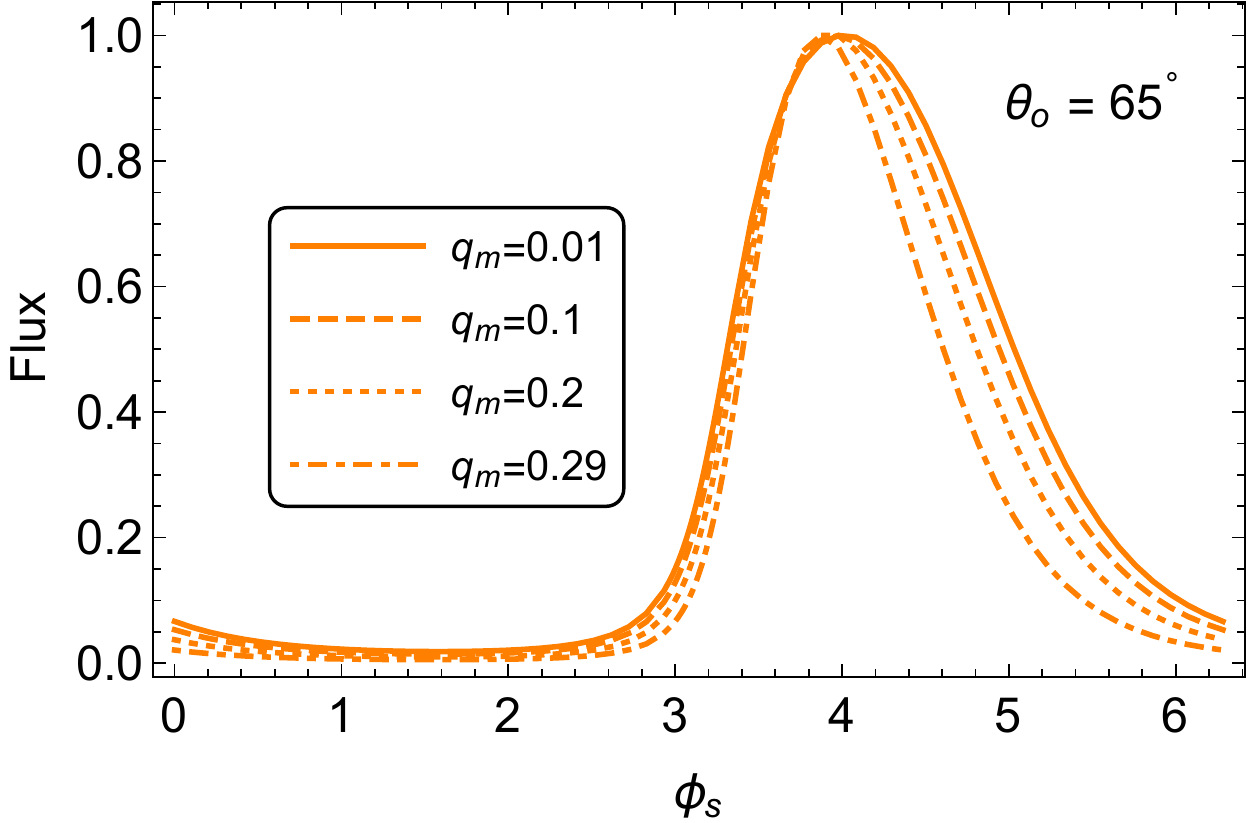}\\
			\includegraphics[width=2.5in]{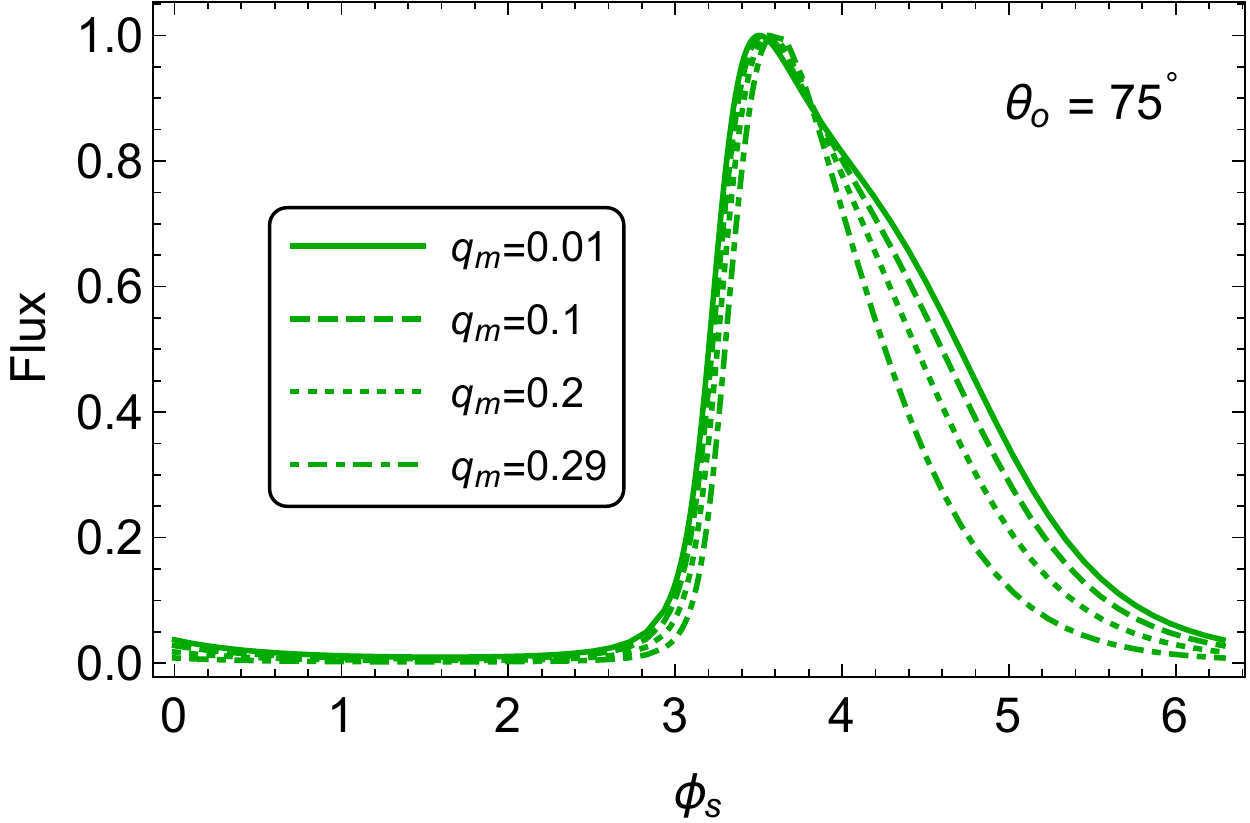}&\includegraphics[width=2.5in]{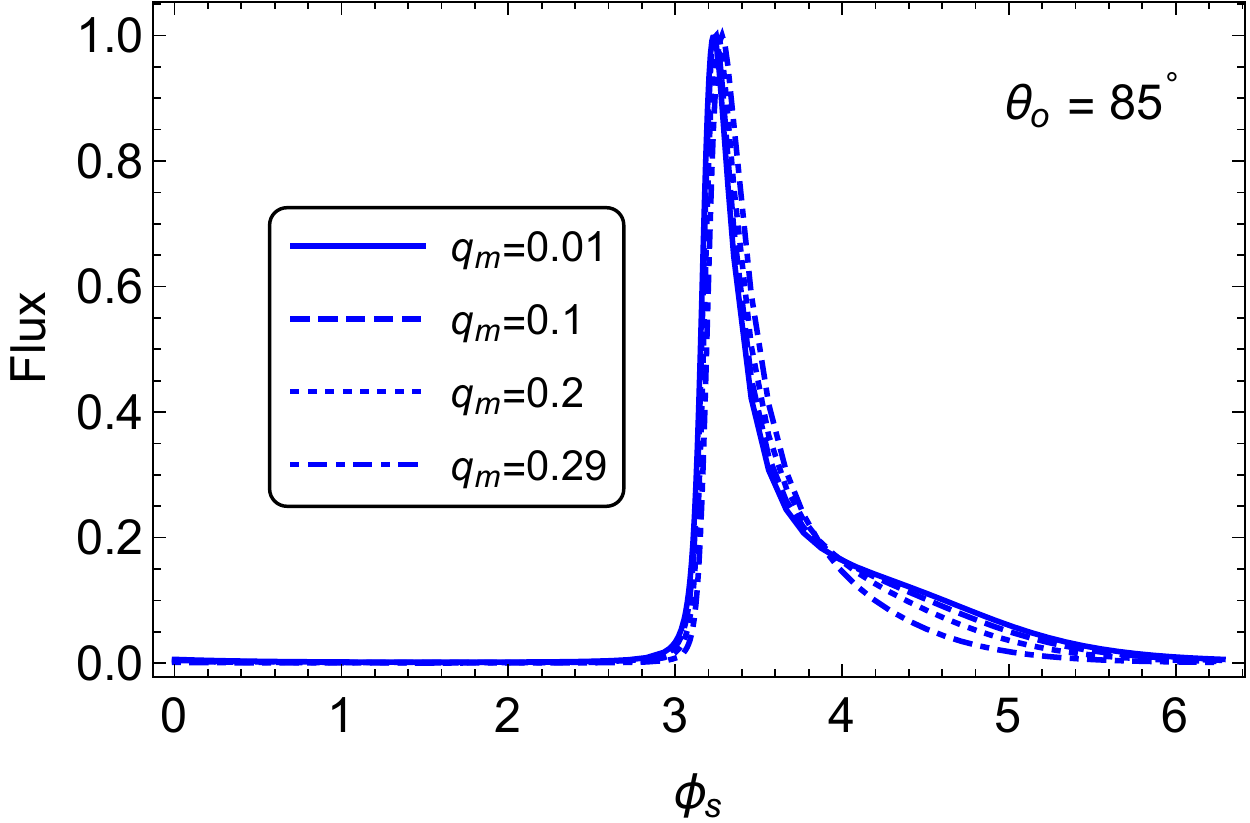}\\
		\end{tabular}
		\caption{The normalized flux as function of hot spots angular position $\phi_s$, for some different values of the inclination angle of the observer $\theta_o$. The spots are located on the ISCOs, which radii depend on the values of the parameter $q_m$. The units for flux are arbitrary.}
		\label{fig:norm_flux}
	\end{center}
\end{figure}

\newpage
The figure ~\ref{fig:norm_flux} demonstrates the normalized flux, the energy measured by a distant observer, as the function of the angular position of the hot spot. The resulting graphs for the flux are a combination of the frequency shift and gravitational lensing. We have normalized it to 1 because it does not make sense to discuss absolute values of the flux if we do not study the nature of the radiation and assume that it is constant and monochromatic. In order to show how  the graph of the flux is qualitatively changed/varied against the angular position of the hot spot which the distant observer can obtain, we have studied the hot spots moving in ISCOs, for four different inclination angles of the observer $\theta_o = 55^{\circ}, 65^{\circ}, 75^{\circ}, 75^{\circ}$ and for four representative values of the charge parameter $q_m = 0.01, 0.1, 0.2, 0.29$. The inclination angle influences the shape of the graph. With increasing the inclination angle, the effect of focusing increases, and the peak on the graph shifts towards the value of the angular position $\phi_s = \pi$. The charge parameter $q_m$ also influences the shape of the graphs, making the peak more narrow for the higher values of the parameter $q_m$. The table ~\ref{tab:table2} presents the numerical values corresponding to the maxima and minima of the graphs.         

\begin{table}[H]
\caption{\label{tab:table2}%
Columns present the inclination angle of the observer $\theta_o$, the charge parameter $q_m$, the radius of the hot spot circular orbit $r_s$, the hot spots angular positions $\phi_s$ corresponding to the maximum and minimum of the flux and the normalized value of the flux at maximum and minimum.
}
\begin{ruledtabular}
\begin{tabular}{ccccccc}
\textrm{$\theta_o$}&
\textrm{$q_m$}&
\textrm{$r_s$}&
\textrm{$\phi_s$}&
\textrm{$MAX(Flux)$}&
\textrm{$\phi_s$}&
\textrm{$MIN(Flux)$}\\
\colrule
 55 & 0.01 & 5.90968 & 4.21249 & 1.0 & 1.56859 & 0.0311783 \\
 55 & 0.1 & 5.06204 & 4.17534 & 1.0 & 1.56393 & 0.0258099 \\
 55 & 0.2 & 4.00497 & 4.12429 & 1.0 & 1.55517 & 0.0194316 \\
 55 & 0.29 & 2.78485 & 4.05527 & 1.0 & 1.53623 & 0.0128306 \\
\colrule
 65 & 0.01 & 5.90968 & 4.01118 & 1.0 & 1.58457 & 0.0184402 \\
 65 & 0.1 & 5.06204 & 3.96794 & 1.0 & 1.58076 & 0.0143876 \\
 65 & 0.2 & 4.00497 & 3.92613 & 1.0 & 1.57324 & 0.00984282 \\
 65 & 0.29 & 2.78485 & 3.87548 & 1.0 & 1.55611 & 0.00548648 \\
\colrule
 75 & 0.01 & 5.90968 & 3.50478 & 1.0 & 1.59589 & 0.00954956 \\
 75 & 0.1 & 5.06204 & 3.52974 & 1.0 & 1.59266 & 0.00702237 \\
 75 & 0.2 & 4.00497 & 3.56297 & 1.0 & 1.58602 & 0.00433415 \\
 75 & 0.29 & 2.78485 & 3.60098 & 1.0 & 1.5702 & 0.00201683 \\
\colrule
 85 & 0.01 & 5.90968 & 3.2311 & 1.0 & 1.60179 & 0.00148454 \\
 85 & 0.1 & 5.06204 & 3.23953 & 1.0 & 1.59886 & 0.00108087 \\
 85 & 0.2 & 4.00497 & 3.25374 & 1.0 & 1.59267 & 0.000645077 \\
 85 & 0.29 & 2.78485 & 3.27907 & 1.0 & 1.57753 & 0.000266911 \\
\end{tabular}
\end{ruledtabular}
\end{table}

\begin{figure}[H]
	\begin{center}
		\begin{tabular}{cc}
			\includegraphics[width=2.5in]{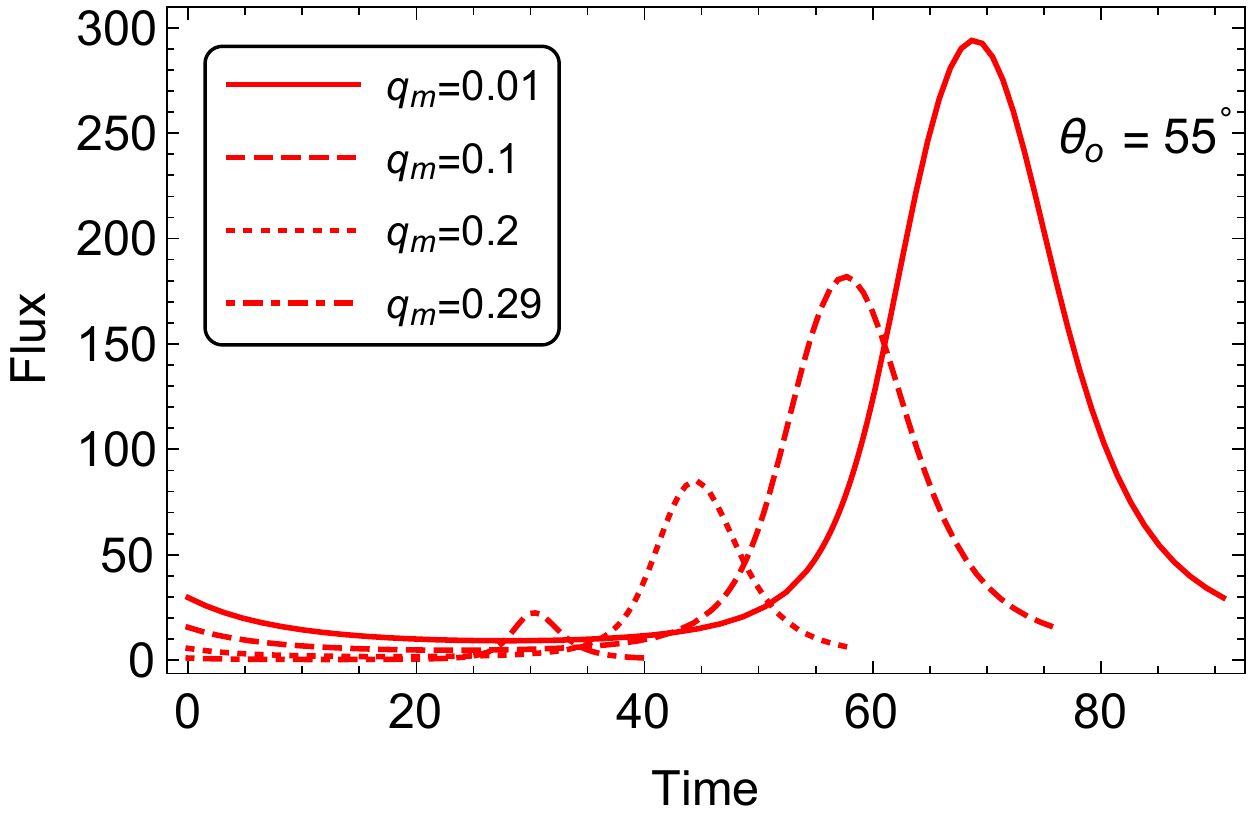}&\includegraphics[width=2.5in]{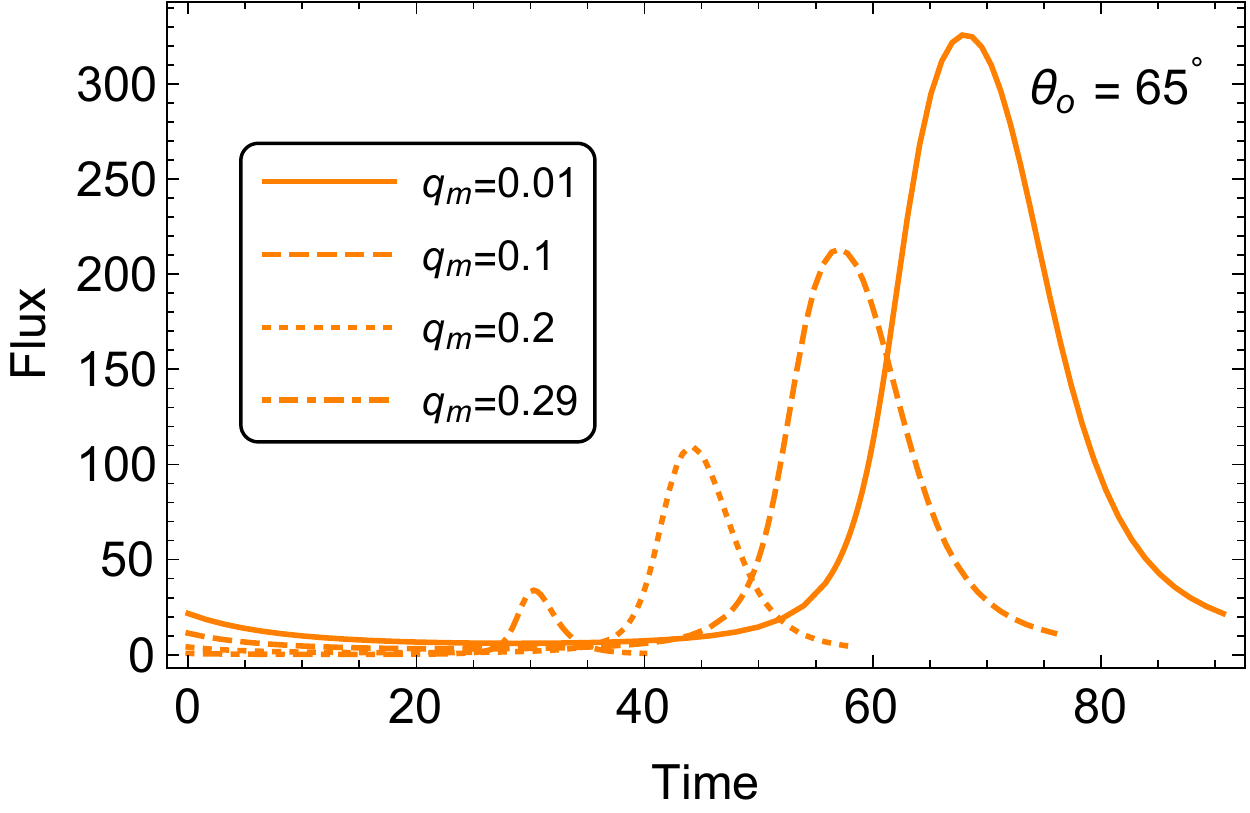}\\
			\includegraphics[width=2.5in]{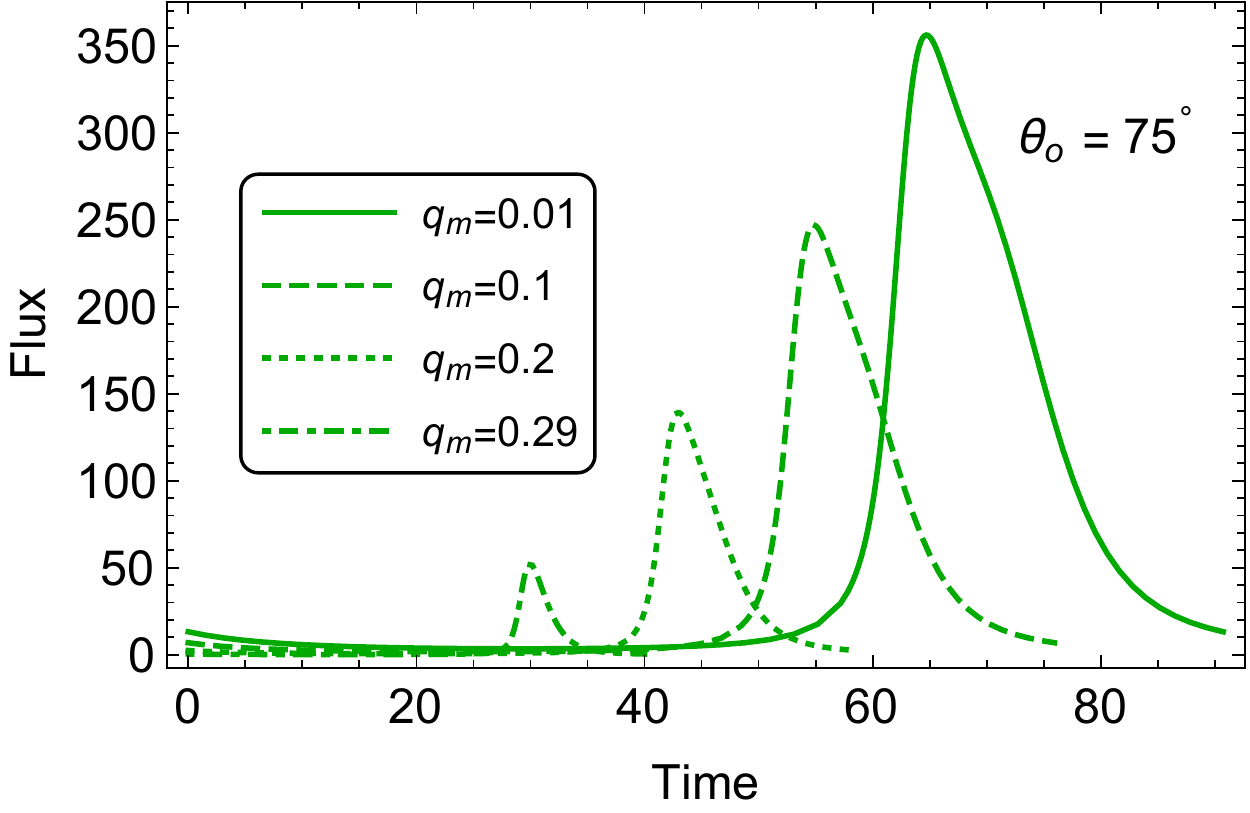}&\includegraphics[width=2.5in]{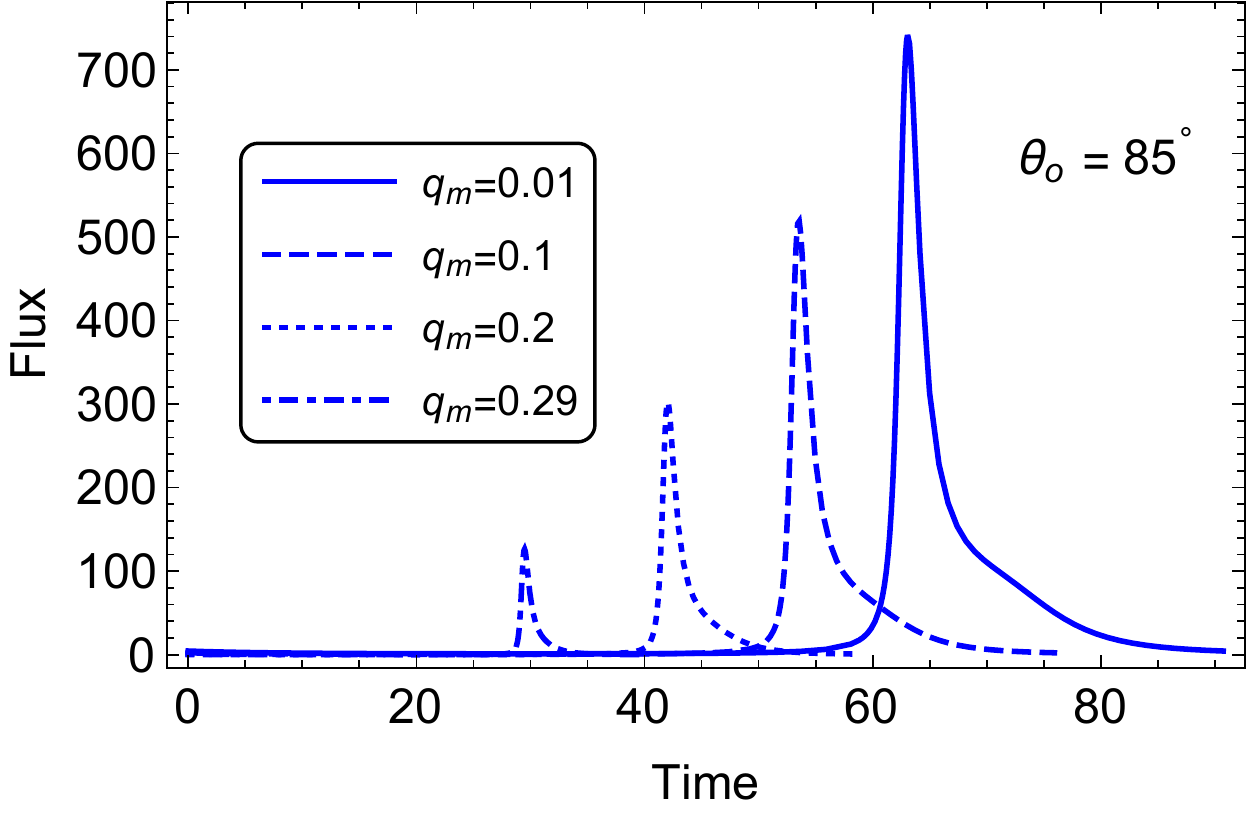}\\
		\end{tabular}
		\caption{The light curves of the hot spots for some different values of the inclination angle of the observer $\theta_o$.The spots are located on the ISCOs, which radii depend on the values of the parameter $q_m$. The horizontal axes shows photons arrival time in the units of $M$. The units for flux are arbitrary.}
		\label{fig:curve}
	\end{center}
\end{figure}

\newpage
The figure ~\ref{fig:curve} demonstrates the light curve, the energy flux measured by the distant observer as the function of the arrival time. We have constructed the light curve of the hot spots on ISCOs for four representative values of the charge parameter $q_m = 0.01, 0.1, 0.2, 0.29$ and four different inclination angles of the observer $\theta_o = 55^{\circ}, 65^{\circ}, 75^{\circ}, 75^{\circ}$. Different values of the parameter $q_m$ correspond to different values of the orbital periods of hot spots. One can notice that for different inclination angles of the observer, the shape of the light curve is also significantly changed. For the small inclination angles, the main effect on the shape of the curve is influenced by the frequency shift, while with increasing the inclination angle, focusing becomes the dominant factor influencing the shape of the light curve, although the influence of the Doppler effect is also enhanced with increase of the inclination angle. The table ~\ref{tab:table3} presents the numerical values corresponding to the maxima and minima of the graphs.

\begin{table}[H]
\caption{\label{tab:table3}%
Rows show the inclination angle of the observer $\theta_o$, the charge parameter $q_m$, the radius of the hot spot circular orbit $r_s$, the observation time $t$ corresponding to the maximum and minimum of the flux and the orbital period of the hot spot $T$.
}
\begin{ruledtabular}
\begin{tabular}{cccccc}
\textrm{$\theta_o$}&
\textrm{$q_m$}&
\textrm{$r_s$}&
\textrm{$MAX(t)$}&
\textrm{$MIN(t)$}&
\textrm{$T$}\\
\colrule
 55 & 0.01 & 5.90968 & 68.888 & 27.8399 & 90.7257 \\
 55 & 0.1 & 5.06204 & 57.6274 & 23.3563 & 75.9301 \\
 55 & 0.2 & 4.00497 & 44.398 & 18.0186 & 58.5133 \\
 55 & 0.29 & 2.78485 & 30.3412 & 12.2416 & 40.0073 \\
\colrule
 65 & 0.01 & 5.90968 & 68.1053 & 28.6905 & 90.7257 \\
 65 & 0.1 & 5.06204 & 57.0293 & 24.0944 & 75.9301 \\
 65 & 0.2 & 4.00497 & 44.0951 & 18.6138 & 58.5133 \\
 65 & 0.29 & 2.78485 & 30.2996 & 12.6679 & 40.0073 \\
\colrule
 75 & 0.01 & 5.90968 & 64.7082 & 29.2891 & 90.7257 \\
 75 & 0.1 & 5.06204 & 54.8434 & 24.6143 & 75.9301 \\
 75 & 0.2 & 4.00497 & 42.9785 & 19.0338 & 58.5133 \\
 75 & 0.29 & 2.78485 & 29.9715 & 12.9694 & 40.0073 \\
\colrule
 85 & 0.01 & 5.90968 & 63.081 & 29.5992 & 90.7257 \\
 85 & 0.1 & 5.06204 & 53.5073 & 24.8836 & 75.9301 \\
 85 & 0.2 & 4.00497 & 42.0305 & 19.2515 & 58.5133 \\
 85 & 0.29 & 2.78485 & 29.4791 & 13.1261 & 40.0073 \\
\end{tabular}
\end{ruledtabular}
\end{table}

\begin{figure}[H]
	\begin{center}
		\begin{tabular}{cc}
			\includegraphics[width=2.5in]{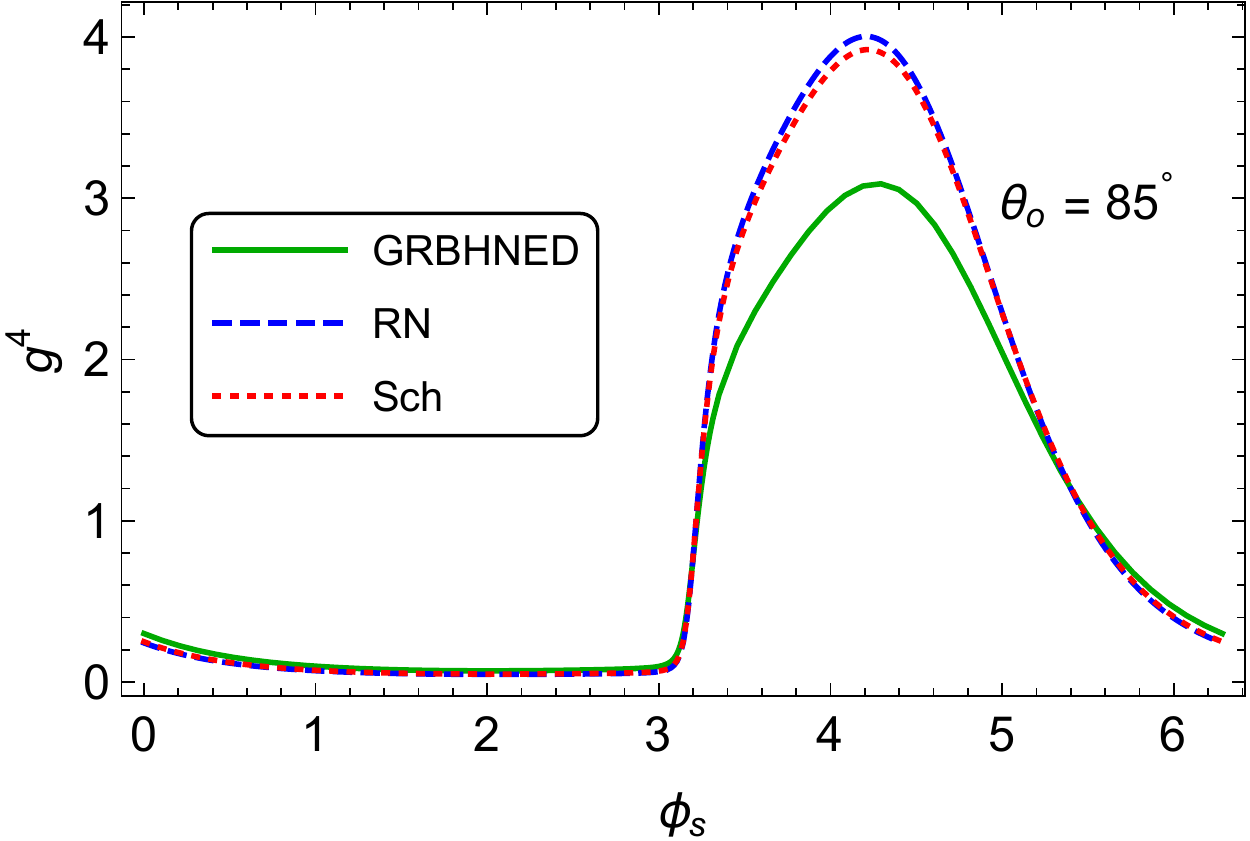}&\includegraphics[width=2.5in]{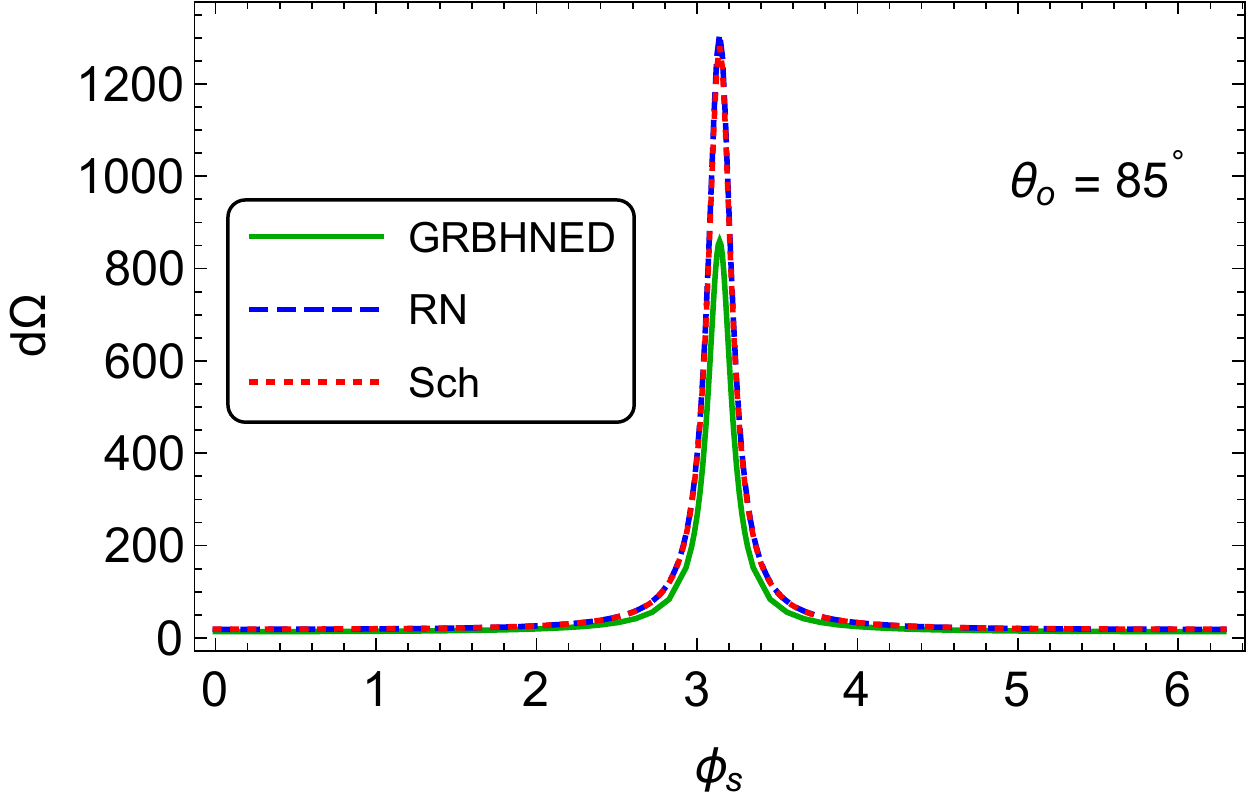}\\
			\includegraphics[width=2.5in]{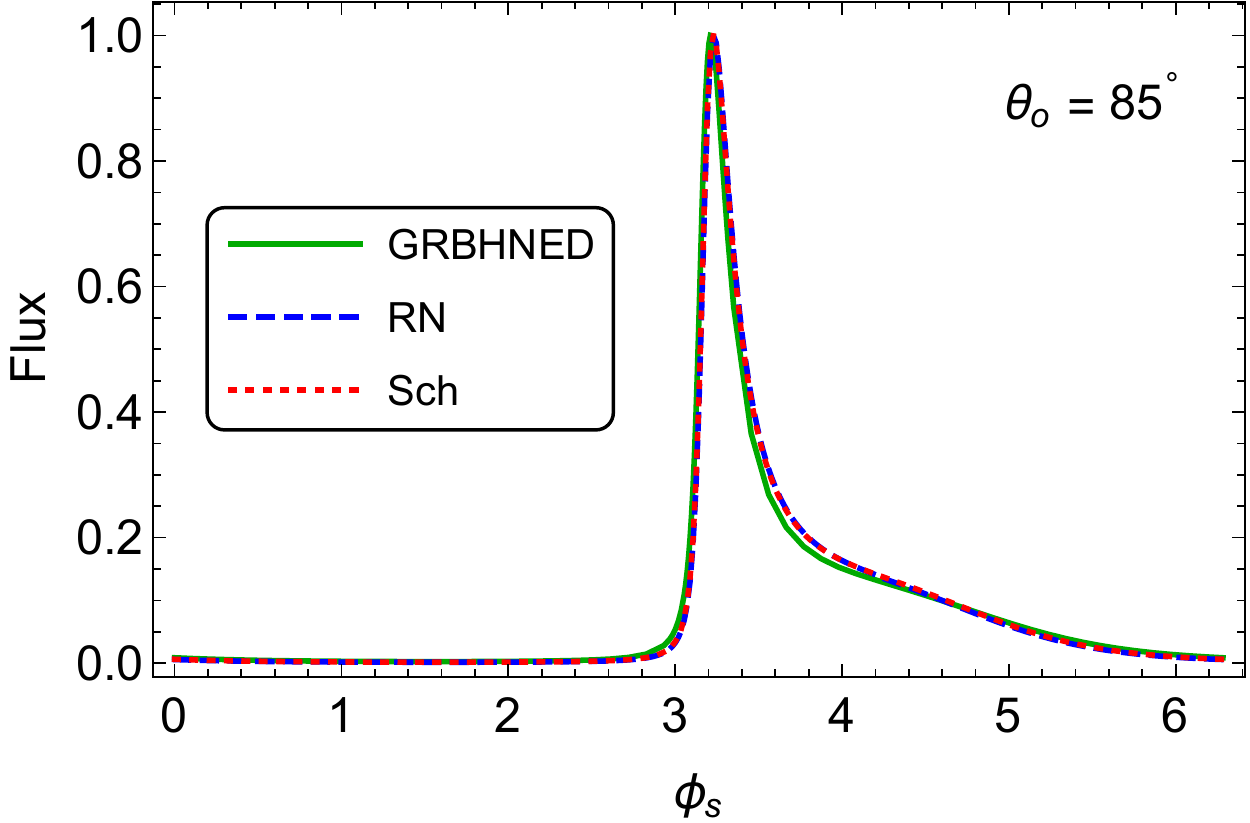}&\includegraphics[width=2.5in]{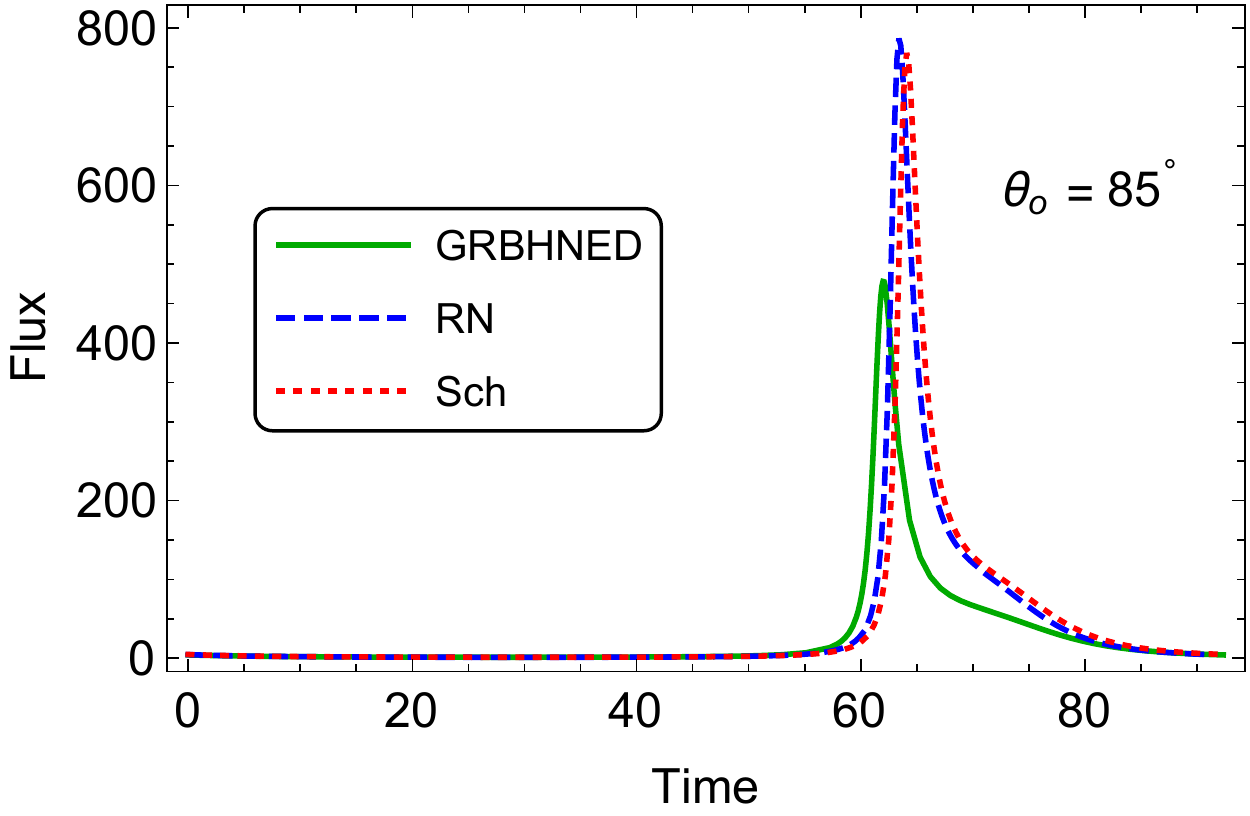}\\
		\end{tabular}
		\caption{The frequency shift $g^4$, the solid angle $\diff\Omega$, the normalized flux as function of hot spots angular position $\phi_s$, and the light curve (the flux as function of time) of the hot spots moving with the same angular velocity around Schwarzschild, Reissner-Nordstrom and Maxwellian regular black holes, for the observer with inclination angle $\theta_o = 85^{\circ}$. The units for frequency shift, solid angle and flux are arbitrary. The arrival time of the photon is in the units of $M$.}
		\label{fig:comp_spacetime}
	\end{center}
\end{figure}

We have compared the frequency shift $g^4$, normalized frequency shift $g^4$, the solid angle $\diff\Omega$, the flux and normalized flux as a function of hot spots angular position $\phi_s$, and the light curve (the flux as a function of arrival time) of the hot spots moving with the same angular velocity around Schwarzschild, Reissner-Nordstrom and Maxwellian regular black hole, for the observer with inclination angle $\theta_o = 85^{\circ}$. In all three cases, the orbit of the hot spot is chosen to keep the orbital period unchanged. The charge parameter is selectted as $q_m = 0.29$ in the case of Reissner-Nordstrom and Maxwellian regular black hole. The figure ~\ref{fig:comp_spacetime} shows the frequency shift $g^4$, the solid angle $\diff\Omega$, the normalized flux as a function of hot spots angular position $\phi_s$, and the light curve (the flux as a function of arrival time). The frequency shift is less substantial in the case of the Maxwellian regular black hole than in the case of Schwarzschild and Reissner-Nordstrom black holes. One can notice that the profile of the normalized flux in the case of the Maxwellian regular black hole slightly differs from Schwarzschild and Reissner-Nordstroms black holes. Although the hot spots are moving along the orbits with the same orbital period, the peak on the light curve that corresponds to the maximum of energy measured by the distant observer is slightly shifted towards smaller values of observational time in case of the Maxwellian regular black hole as compared with the Schwarzschild, Reissner-Nordstrom black holes. The tables ~\ref{tab:table4a}, \ref{tab:table4b} and \ref{tab:table4c} show the numerical values corresponding to the maxima and minima of the graphs.

\newpage
\begin{table}[H]
\caption{\label{tab:table4a}%
Columns demonstrate the radius of the hot spot circular orbit $r_s$, the hot spots angular positions $\phi_s$ corresponding to the maximum and minimum of the frequency shift $g^4$ and the maximum and minimum value of the frequency shift $g^4$. The inclination angle of the observer is $\theta_o = 85^{\circ}$. 
}
\begin{ruledtabular}
\begin{tabular}{ccccc}
\textrm{$r_s$}&
\textrm{$\phi_s$}&
\textrm{$MAX(g^4)$}&
\textrm{$\phi_s$}&
\textrm{$MIN(g^4)$}\\
\colrule
 5.38632 & 4.27171 & 3.09089 & 2.01154 & 0.0696881 \\
 5.9717 & 4.20787 & 4.00361 & 2.07532 & 0.0475264 \\
 6. & 4.21766 & 3.9219 & 2.06552 & 0.0497094 \\
\end{tabular}
\end{ruledtabular}
\end{table}
\begin{table}[H]
\caption{\label{tab:table4b}%
Columns present the radius of the hot spot circular orbit $r_s$, the hot spots angular positions $\phi_s$ corresponding to the maximum and minimum of the flux and the normalized value of the flux at maximum and minimum. The inclination angle of the observer is $\theta_o = 85^{\circ}$.
}
\begin{ruledtabular}
\begin{tabular}{ccccc}
\textrm{$r_s$}&
\textrm{$\phi_s$}&
\textrm{$MAX(Flux)$}&
\textrm{$\phi_s$}&
\textrm{$MIN(Flux)$}\\
\colrule
 5.38632 & 3.21514 & 1.0 & 1.55716 & 0.00249092 \\
 5.9717 & 3.23205 & 1.0 & 1.60502 & 0.00141309 \\
 6. & 3.23032 & 1.0 & 1.60199 & 0.00153013 \\
\end{tabular}
\end{ruledtabular}
\end{table}
\begin{table}[H]
\caption{\label{tab:table4c}%
Columns present the radius of the hot spot circular orbit $r_s$, the observation time $t$ corresponding to the maximum and minimum of the flux, the orbital period of the hot spot $T$ and the difference in observational time for the maximum and minimum of the flux. The inclination angle of the observer is $\theta_o = 85^{\circ}$.
}
\begin{ruledtabular}
\begin{tabular}{ccccc}
\textrm{$r_s$}&
\textrm{$MAX(t)$}&
\textrm{$MIN(t)$}&
\textrm{$T$}&
\textrm{$MAX(t)-MIN(t)$}\\
\colrule
 5.38632 & 62.0321 & 28.4928 & 92.3436 & 33.5393 \\
 5.9717 & 63.4403 & 29.8195 & 91.0521 & 33.6208 \\
 6. & 64.1198 & 30.1107 & 92.3436 & 34.0091 \\
\end{tabular}
\end{ruledtabular}
\end{table}

\begin{figure}[H]
	\begin{center}
		\begin{tabular}{ccc}
			\includegraphics[width=2.5in]{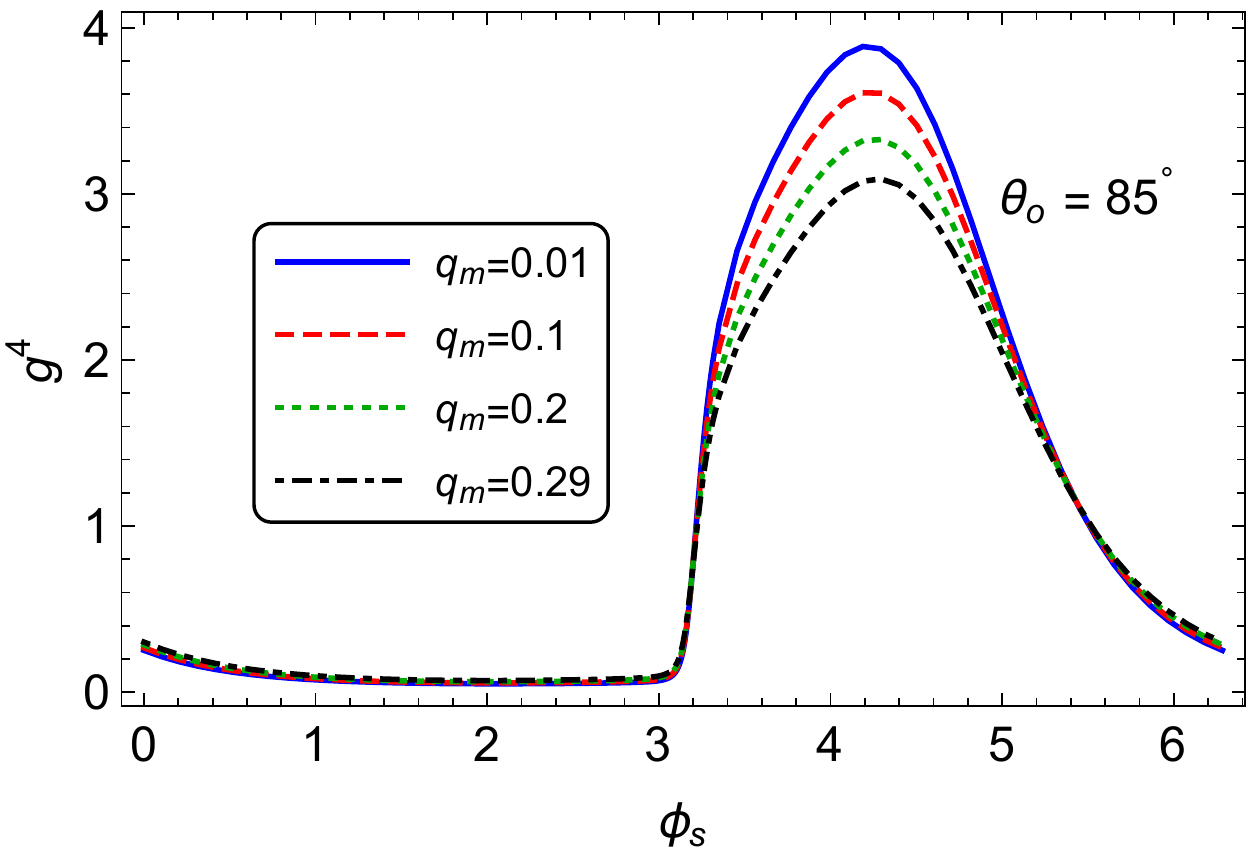}&\includegraphics[width=2.5in]{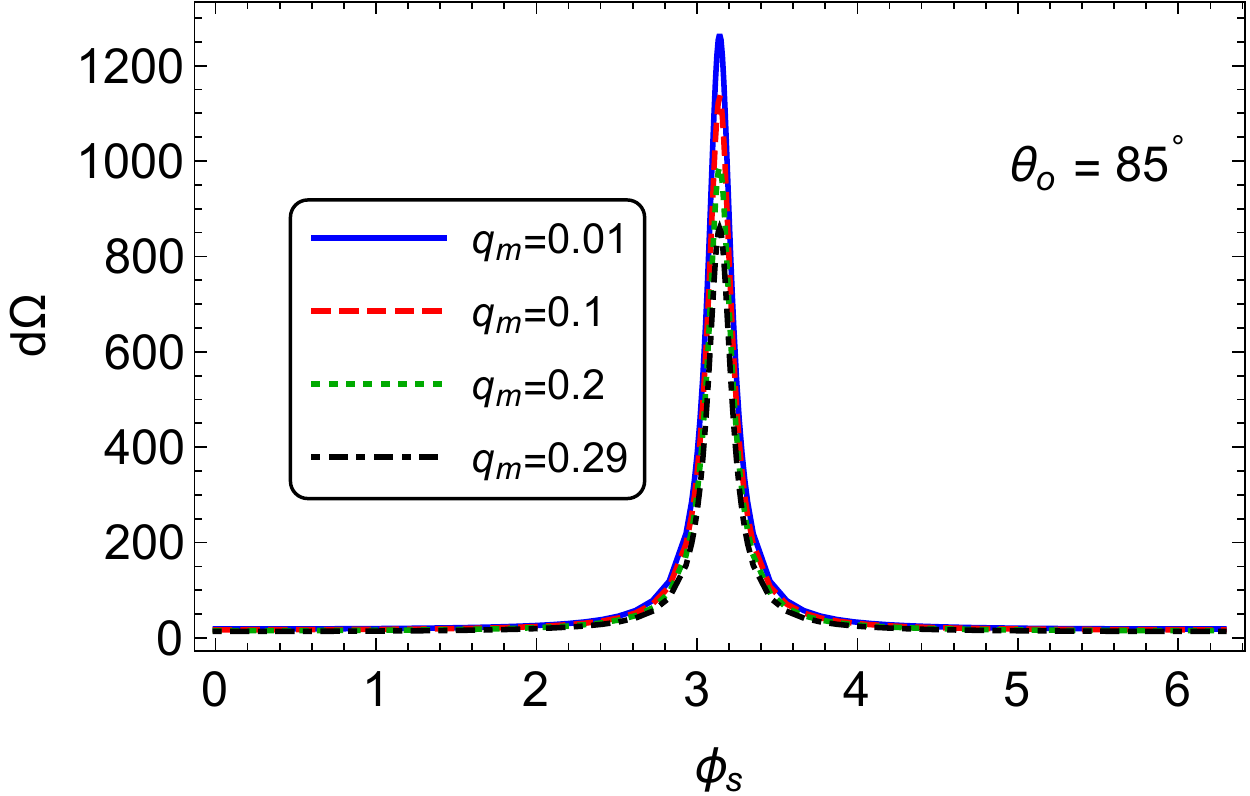}\\
			\includegraphics[width=2.5in]{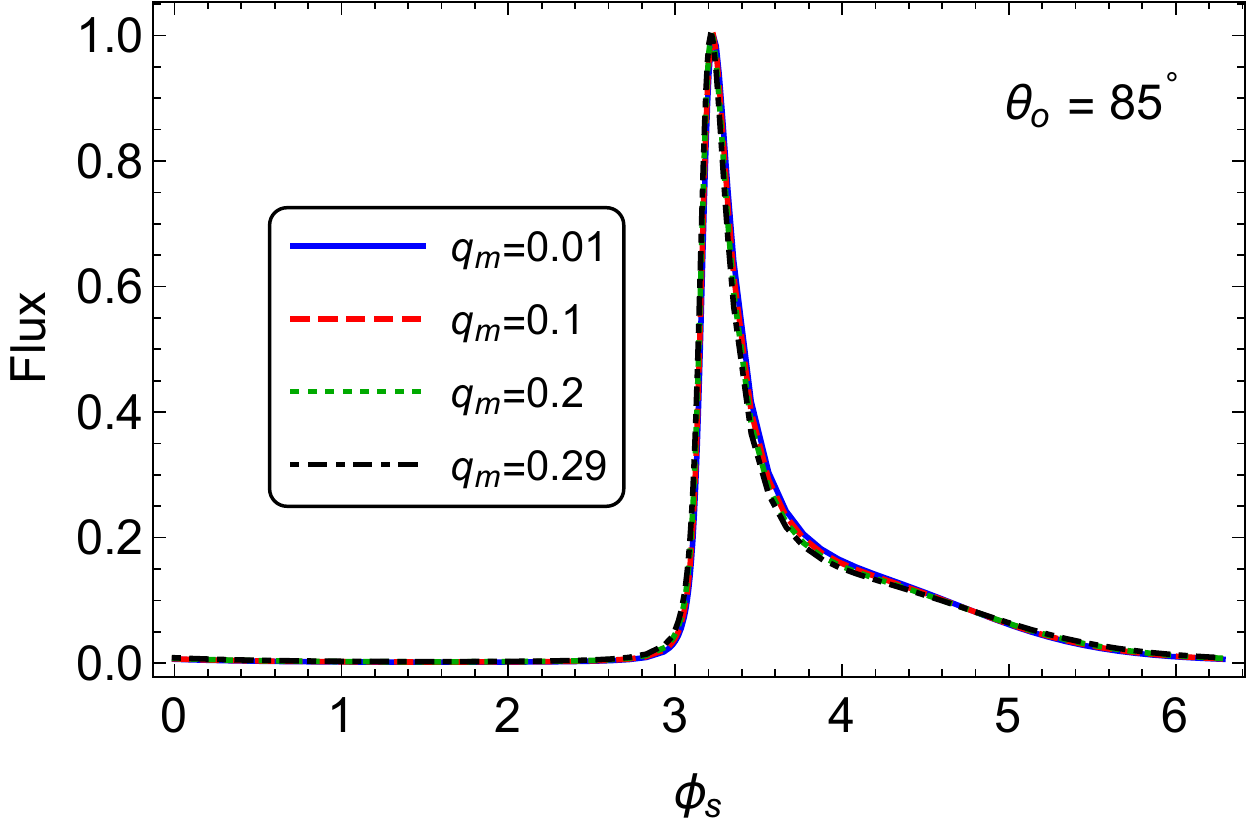}&\includegraphics[width=2.5in]{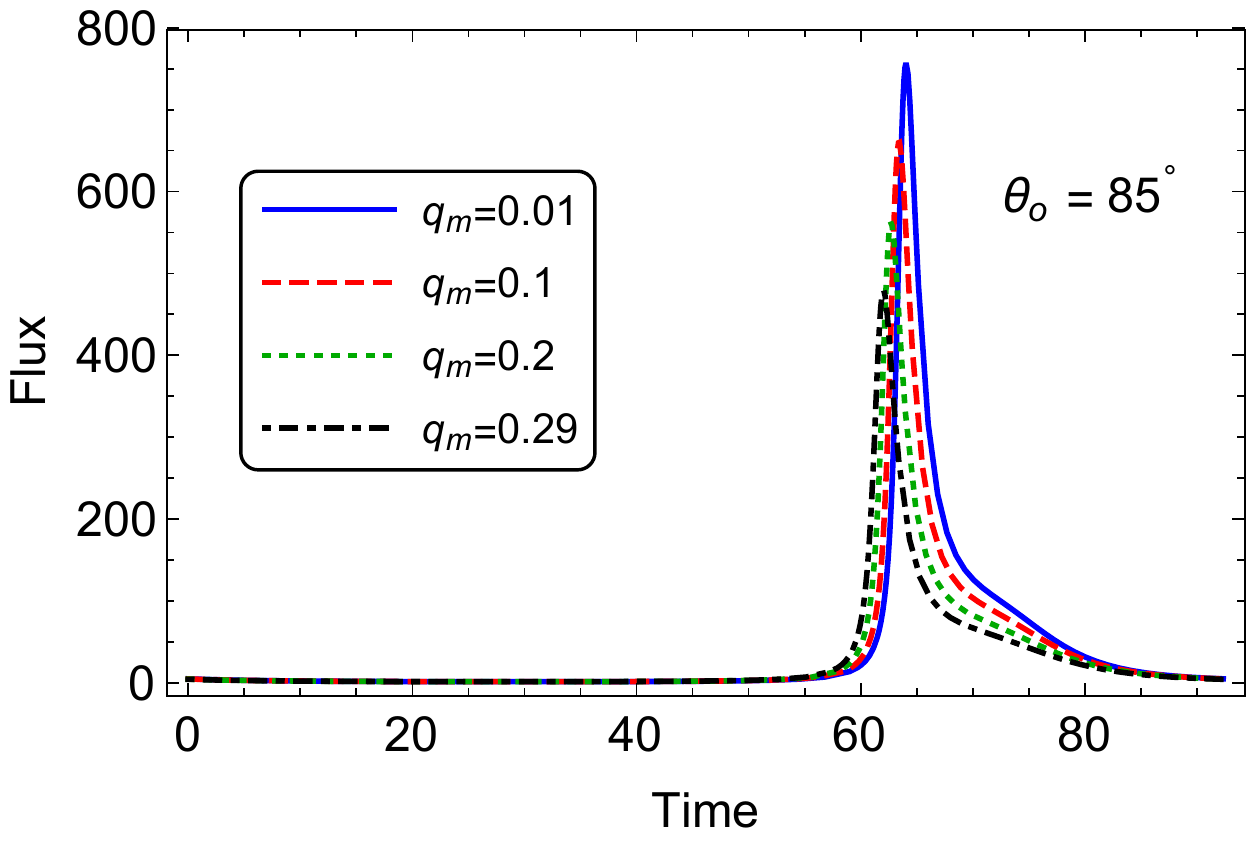}\\
		\end{tabular}
		\caption{Comparison of the frequency shift $g^4$, the solid angle $\diff\Omega$, the normalized flux as function of hot spots angular position $\phi_s$, and the light curve (the flux as function of time) of the hot spots moving with the same angular velocity around Maxwellian regular black hole for several representative values of the parameter $q_m$, for the observer with inclination angle $\theta_o = 85^{\circ}$. The units for frequency shift, solid angle and flux are arbitrary. The arrival time of the photon is in the units of $M$.}
		\label{fig:sav_spacetime}
	\end{center}
\end{figure}

\newpage
The figure ~\ref{fig:sav_spacetime} shows the comparison of the frequency shift $g^4$, the solid angle $\diff\Omega$, the normalized flux as a function of hot spots angular position $\phi_s$, and the light curve (the flux as a function of arrival time) for hot spots rotating around the Maxwellian regular black hole for four representative values of the charge parameter $q_m$. The radii of the orbits are chosen to keep the orbital periods of the hot spots equal in all four cases. The frequency shift is decreased with increasing the charge parameter $q_m$. The peaks on the plot, where the normalized flux is compared, with increasing the charge parameter $q_m$ are shifted towards the value of the angular position of the hot spot at $\phi = \pi$. The peaks on the light curves plot are also shifted towards the smaller arrival time values with increasing the charge parameter $q_m$. The tables ~\ref{tab:table5a}, \ref{tab:table5b} and \ref{tab:table5c} show the numerical values corresponding to the maxima and minima of the graphs.

\begin{table}[H]
\caption{\label{tab:table5a}%
Columns show the charge parameter $q_m$, the radius of the hot spot circular orbit $r_s$, the hot spots angular positions $\phi_s$ corresponding to the maximum and minimum of the frequency shift $g^4$ and the maximum and minimum value of the frequency shift $g^4$. The inclination angle of the observer is $\theta_o = 85^{\circ}$. 
}
\begin{ruledtabular}
\begin{tabular}{cccccc}
\textrm{$q_m$}&
\textrm{$r_s$}&
\textrm{$\phi_s$}&
\textrm{$MAX(g^4)$}&
\textrm{$\phi_s$}&
\textrm{$MIN(g^4)$}\\
\colrule
 0.01 & 5.97997 & 4.21954 & 3.89009 & 2.06358 & 0.0502911 \\
 0.1 & 5.79647 & 4.23674 & 3.61505 & 2.04644 & 0.0558334 \\
 0.2 & 5.58495 & 4.2554 & 3.33063 & 2.02783 & 0.0627164 \\
 0.29 & 5.38632 & 4.27171 & 3.09089 & 2.01154 & 0.0696881 \\
\end{tabular}
\end{ruledtabular}
\end{table}
\begin{table}[H]
\caption{\label{tab:table5b}%
Columns present the charge parameter $q_m$, the radius of the hot spot circular orbit $r_s$, the hot spots angular positions $\phi_s$ corresponding to the maximum and minimum of the flux and the normalized value of the flux at maximum and minimum. The inclination angle of the observer is $\theta_o = 85^{\circ}$.
}
\begin{ruledtabular}
\begin{tabular}{cccccc}
\textrm{$q_m$}&
\textrm{$r_s$}&
\textrm{$\phi_s$}&
\textrm{$MAX(Flux)$}&
\textrm{$\phi_s$}&
\textrm{$MIN(Flux)$}\\
\colrule
 0.01 & 5.97997 & 3.22978 & 1.0 & 1.60057 & 0.0015549 \\
 0.1 & 5.79647 & 3.22504 & 1.0 & 1.58721 & 0.00180259 \\
 0.2 & 5.58495 & 3.21983 & 1.0 & 1.57174 & 0.00212878 \\
 0.29 & 5.38632 & 3.21514 & 1.0 & 1.55716 & 0.00249092 \\
\end{tabular}
\end{ruledtabular}
\end{table}
\begin{table}[H]
\caption{\label{tab:table5c}%
Shows the charge parameter $q_m$, the radius of the hot spot circular orbit $r_s$, the observation time $t$ corresponding to the maximum and minimum of the flux, the orbital period of the hot spot $T$ and the difference in observational time for the maximum and minimum of the flux. The inclination angle of the observer is $\theta_o = 85^{\circ}$.
}
\begin{ruledtabular}
\begin{tabular}{cccccc}
\textrm{$q_m$}&
\textrm{$r_s$}&
\textrm{$MAX(t)$}&
\textrm{$MIN(t)$}&
\textrm{$T$}&
\textrm{$MAX(t)-MIN(t)$}\\
\colrule
 0.01 & 5.97997 & 64.0511 & 30.0579 & 92.3436 & 33.9932 \\
 0.1 & 5.79647 & 63.4228 & 29.5681 & 92.3436 & 33.8548 \\
 0.2 & 5.58495 & 62.7031 & 29.0098 & 92.3436 & 33.6932 \\
 0.29 & 5.38632 & 62.0321 & 28.4928 & 92.3436 & 33.5393 \\
\end{tabular}
\end{ruledtabular}
\end{table}

\begin{figure}[H]
	\begin{center}
		\begin{tabular}{ccc}
			\includegraphics[width=2.5in]{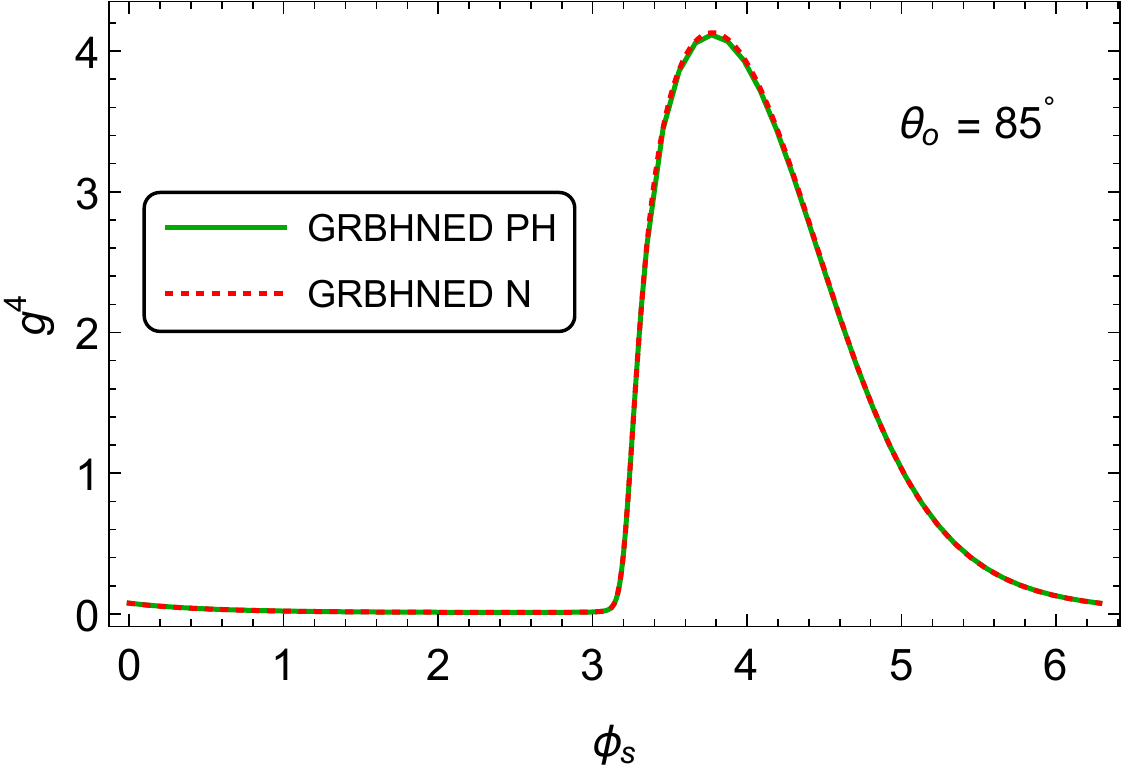}&\includegraphics[width=2.5in]{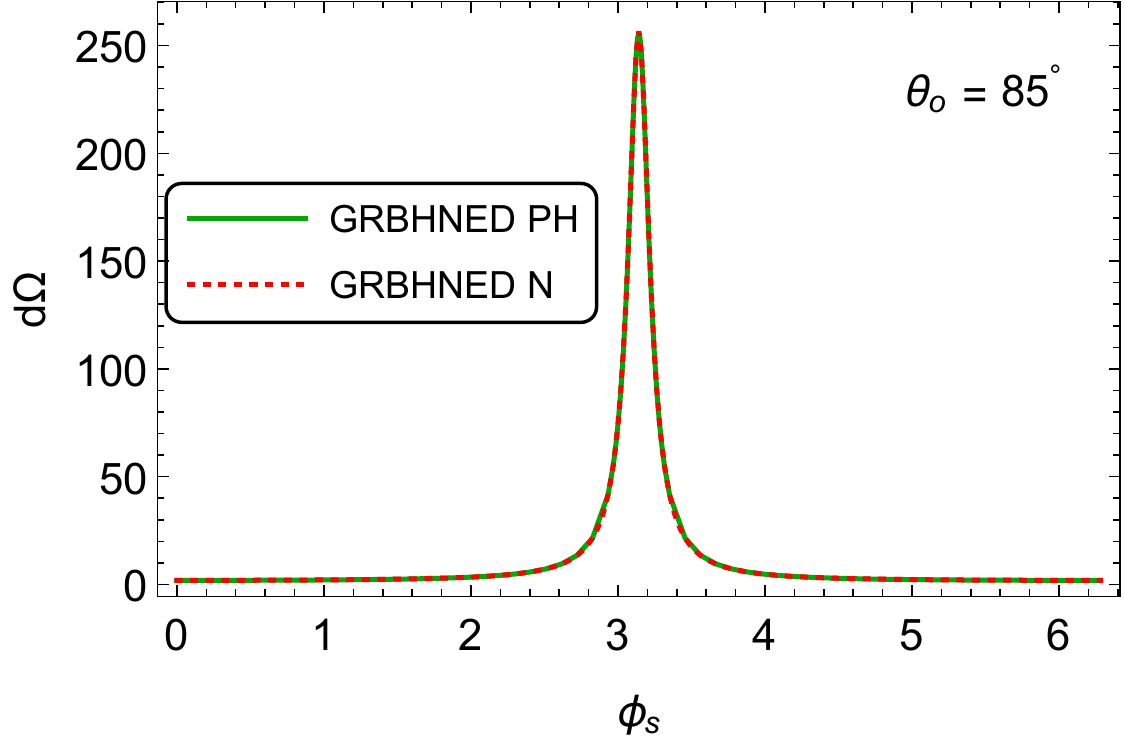}\\
			\includegraphics[width=2.5in]{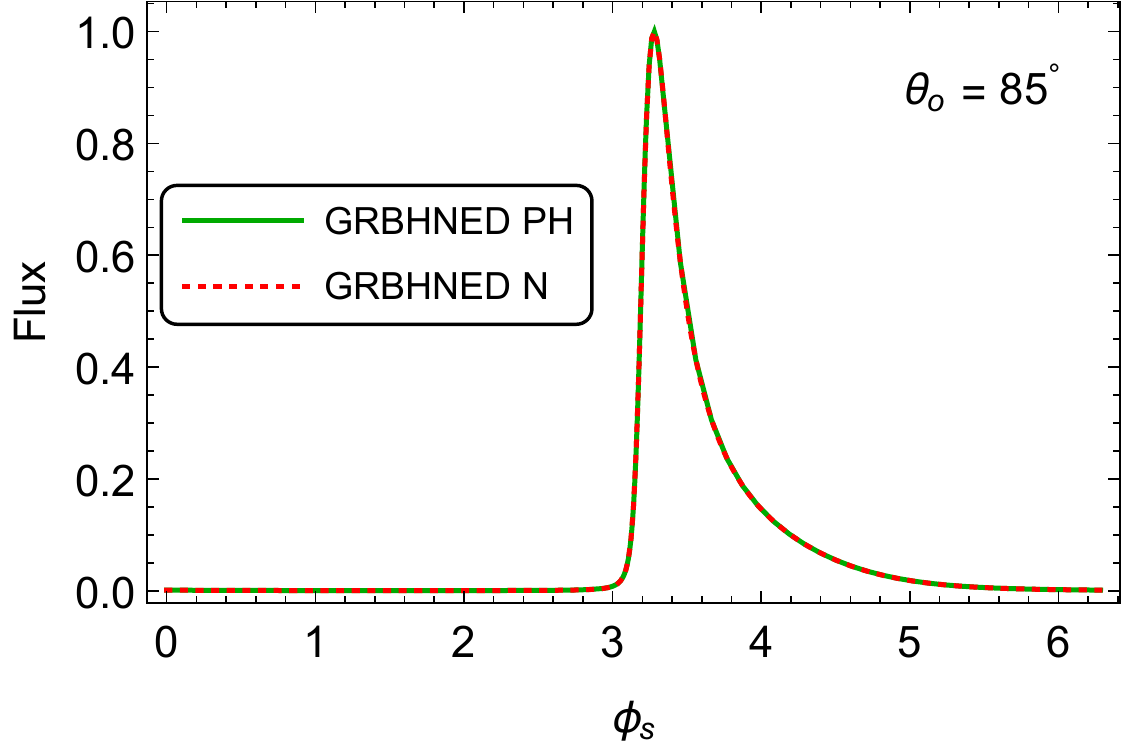}&\includegraphics[width=2.5in]{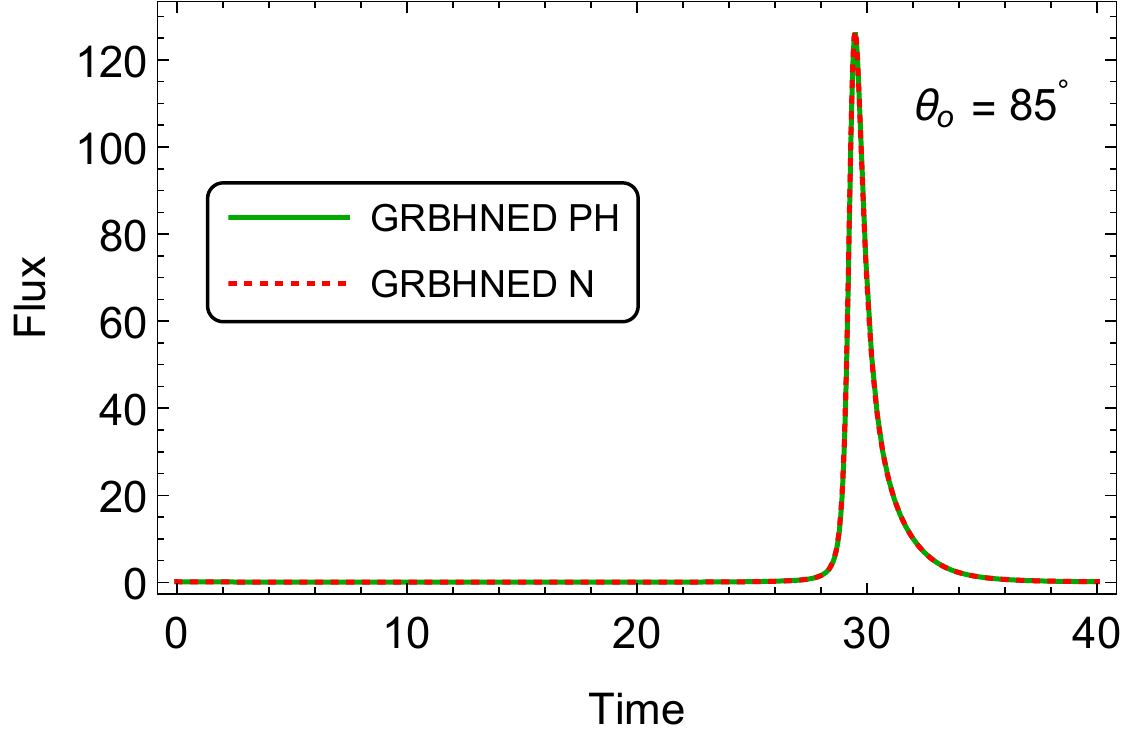}\\
		\end{tabular}
		\caption{The frequency shift $g^4$, the solid angle $\diff\Omega$, the normalized flux as function of hot spots angular position $\phi_s$, and the light curve (the flux as function of time) of the hot spot moving on ISCO around Maxwellian regular black hole for the charge parameter $q_m = 0.29$, for the observer with inclination angle $\theta_o = 85^{\circ}$. The units for frequency shift, solid angle and flux are arbitrary. The arrival time of the photon is in the units of $M$. The graphs of the photons moving in effective geometry compared with graph of the neutrinos moving in the spacetime geometry.}
		\label{fig:neut_ph_comp}
	\end{center}
\end{figure}

\newpage
The figure ~\ref{fig:neut_ph_comp} shows the comparison of the frequency shift $g^4$, the solid angle $\diff\Omega$, the normalized flux as a function of hot spots angular position $\phi_s$, and the light curve (the flux as a function of arrival time) for the photons and neutrinos radiated by the hot spots rotating around the Maxwellian regular black hole for four representative values of the charge parameter $q_m$. In both cases, the hot spots are located on ISCOs. The graphs of the frequency shift, the graph of the hotspots solid angle at which it is visible, normalized flux and peak on the light curve almost coincide for neutrinos moving in spacetime geometry and photons moving in effective geometry of the Maxwellian regular black hole.

\section{\label{sec:dis-concl}Discussion and conclusion}

The main aim of the performed  research was to analyze fingerprints of Maxwellian regular black hole in light curve of point-like hot-spots orbiting  on Keplerian circular orbit. 

We have constructed four kind of plots: (1) the frequency shift $g^4$ as function of hot-spot azimuthal angle $\phi_S$ carrying information about Doppler and gravitational changes of photon energy, (2) the solid angle $\diff\Omega$ as function of $\phi_S$, represents the gravitational lensing effects, (3) total observed flux as function of $\phi_S$ and (4) the total observed flux as function of time of arrival.

We have simulated light curves for four representative values of charge parameter $q_m=0.01$, $0.1$, $0.2$, and $0.29$ and for four representative values of observer latitude $\theta_o=55^\circ$, $65^\circ$, $75^\circ$, and $85^\circ$.

In order to extract the pure effect of Maxwellian effective geometry on light curve of the hotspot orbiting central black hole we selected the circular orbits in specific way that the angular frequency is kept fixed and we modify the $q_m$ parameter (selecting representative values of $q_m=0.01$, $0.1$, $0.2$, and $0.29$) and accordingly construct quantities $g^4$, $\diff\Omega$, $F_{tot}$ describing light curve.

We have found that the value of $q_m$ is imprinted in the magnitude of the frequency shift curve $g(t_{obs})$ maximum, the smaller is the value of $q_m$ the higher is the value of $g^4$ maximum. On the other hand, comparing shapes of light curves ($F_{tot}=F_{tot}(t_{obs})$) we do not observe any significant influence of $q_m$ parameter on it.

We have also compared effect of effective geometry on light curve when compared to the case of radiation propagation along null geodesics of background metric. We have found that in the case of Maxwellian regular black hole the effect is very small. 

There is an interesting observation of light curve maximum shift, it is due to different values of time delay $\Delta t_o$ of radiation that comes from different orbits which depend of value of $q_m$ (keeping $\Omega$ fixed, it means that orbits are different for different values of parameter $q_m$). The value of time delay is smallest for highest value of parameter $q_m$ and it corresponds to observation that the highest shift of light curve maximum is for smallest value of $q_m$. Similar behavior is also observed in case of hot spots located on ISCO.

From the performed study we have learnt the period of the hot-spot orbit can be directly read-off from the light curve period. By using and applying the information from the dependence $g^4_{max}(q_m)=\mathrm{MAX}(g^4(q_m))$ one can determine the parameters of the hot spot orbit and the parameters of the  spacetime. We plan to present and develop this procedure in the subsequent paper.

\begin{acknowledgments}
We would like to acknowledge the institutional support of Silesian University in Opava. This work was supported by the Student Grant Foundation of the Silesian University in Opava, Grant No. SGF/3/2020, which was realized within the EU OPSRE project entitled "Improving the quality of the internal grant scheme of the Silesian University in Opava", reg. number: CZ.02.2.69/0.0/0.0/19\_073/0016951. D. O., J. S. and Z. S. also acknowledge the grant 19-03950S of Czech Science Foundation(GACR) and  the internal student grant of the Silesian University in Opava SGS/12/2019. B. A. acknowledges the Uzbekistan Ministry for Innovative Development and the Abdus Salam International Centre for Theoretical Physics for support under the Grant No. OEA NT-01.
\end{acknowledgments}

\appendix

\section{\label{sec:app-1}Derivation of the expression for partial derivative $\partial u_s / \partial b$}

In case when impact parameter $b < b_{ph}$ the partial derivative $\partial u_s / \partial b$ is equal to 
\begin{equation}
\frac{\partial u_s}{\partial b}\bigg\vert_{b < b_{ph}} = -\frac{\partial F_1 / \partial b}{\partial F_1 / \partial u_s} \ ,
\end{equation}
where  $F_1$ is the function defined in (\ref{eq:f1}) ($F_1(u_s, b) = \varphi(\theta_o, \phi_s) - \int_{u_{o}}^{u_{s}} \frac{\diff u}{\sqrt{U(u, b)}}$). Because the first term in the definition of the function $F_1$ does not depend on $b$ and $u_s$, the partial derivative of this function with respect to $b$ reads  
\begin{equation}
\frac{\partial F_1}{\partial b} = \frac{\partial}{\partial b} \int_{u_{o}}^{u_{s}}\frac{\diff u}{\sqrt{U(u, b)}} = - \int_{u_{o}}^{u_{s}}\frac{\partial U / \partial b}{2 U^{3/2}}\diff u\ ,
\end{equation}
and the partial derivative $\partial F_1 / \partial u_s$ is equal to
\begin{equation}
\begin{split}
    \frac{\partial F_1}{\partial u_s} = \frac{\partial}{\partial u_s} \int_{u_{o}}^{u_{s}}\frac{\diff u}{\sqrt{U(u, b)}} &= \lim_{\Delta u_s \to 0} \frac{1}{\Delta u_s} \left( \int_{u_{o}}^{u_{s} + \Delta u_s}\frac{\diff u}{\sqrt{U(u, b)}} - \int_{u_{o}}^{u_{s}}\frac{\diff u}{\sqrt{U(u, b)}} \right) \\
    & = \lim_{\Delta u_s \to 0} \frac{1}{\Delta u_s} \left( \int_{u_{s}}^{u_{s} + \Delta u_s}\frac{\diff u}{\sqrt{U(u, b)}} \right) \\ 
    &= \lim_{\Delta u_s \to 0} \frac{\Delta u_s}{\sqrt{U(u_s, b)} \, \Delta u_s} = \frac{1}{\sqrt{U(u_s, b)}}\ .
\end{split}
\end{equation}

In case when impact parameter $b > b_{ph}$ the partial derivative $\partial u_s / \partial b$ reads
\begin{equation}
\frac{\partial u_s}{\partial b}\bigg\vert_{b > b_{ph}} = -\frac{\partial F_2 / \partial b}{\partial F_2 / \partial u_s},
\end{equation}
where  $F_2$ is the function defined in (\ref{eq:f2}) ($F_2(u_s, b) = \varphi(\theta_o, \phi_s) - \int_{u_{o}}^{u_{t}(b)}\frac{\diff u}{\sqrt{U(u, b)}}-\int_{u_{s}}^{u_{t}(b)}\frac{\diff u}{\sqrt{U(u, b)}}$).

The partial derivative $\partial F_2 / \partial u_s$ can be found in the similar way how it was done for the function $F_1$, which is equal to 
\begin{equation}
\frac{\partial F_2}{\partial u_s} = \frac{\partial}{\partial u_s} \int_{u_{s}}^{u_{t}(b)}\frac{\diff u}{\sqrt{U(u, b)}} = \frac{1}{\sqrt{U(u_s, b)}}.
\end{equation}

In order to find partial derivative $\partial F_2 / \partial b$, firstly using Newton–Leibniz axiom, we introduce the notation
\begin{equation}
\mathcal{F}(x_t(b),b) - \mathcal{F}(x_o,b) = \int_{x_{o}}^{x_{t}(b)} f(x^{\prime},b) \, \diff x^{\prime},
\end{equation}
where $\mathcal{F}(x^{\prime},b)$ is antiderivative for function $f(x^{\prime},b)$. Taking the derivative with respect to $b$ of both sides of this expression, one can obtain
\begin{equation}
\begin{split}
    \frac{\partial}{\partial b} \int_{x_{o}}^{x_{t}(b)} f(x^{\prime},b) \, \diff x^{\prime} &= \frac{\partial}{\partial b} \mathcal{F}(x_t(b),b) - \frac{\partial}{\partial b} \mathcal{F}(x_o,b)\\
    &= \frac{\partial x_t}{\partial b} \frac{\partial}{\partial x_t} \mathcal{F}(x_t(b),b) + \left(\frac{\partial }{\partial b}\mathcal{F}(x,b)\right)\bigg\vert_{x = x_t(b)} - \frac{\partial}{\partial b} \mathcal{F}(x_o,b)\\
    &=\frac{\partial x_t}{\partial b} f(x_{t}(b),b) + \left(\frac{\partial }{\partial b}\left[\mathcal{F}(x,b) - \mathcal{F}(x_o,b)\right]\right)\bigg\vert_{x = x_t(b)} \\
    &= \frac{\partial x_t}{\partial b} f(x_{t}(b),b) + \left( \frac{\partial}{\partial b} \int_{x_{o}}^{x} f(x^{\prime},b) \, \diff x^{\prime} \right)\bigg\vert_{x = x_t(b)} \\
    &= \frac{\partial x_t}{\partial b} f(x_{t}(b),b) + \int_{x_{o}}^{x_{t}(b)} \frac{\partial}{\partial b} \left( f(x^{\prime},b) \right) \, \diff x^{\prime}\ .
\end{split}
\end{equation}

Thus, applying obtained expression, the partial derivative $\partial F_2 / \partial b$ reads
\begin{equation}
\begin{split}
\frac{\partial}{\partial b} F_2(u_s, b) &= \frac{\partial}{\partial b} \left( \varphi(\theta_o, \phi_s) - \int_{u_{o}}^{u_{t}(b)}\frac{\diff u}{\sqrt{U(u, b)}}-\int_{u_{s}}^{u_{t}(b)}\frac{\diff u}{\sqrt{U(u, b)}} \right)\\
&=- \frac{\partial}{\partial b} \int_{u_{o}}^{u_{t}(b)}\frac{\diff u}{\sqrt{U(u, b)}} - \frac{\partial}{\partial b} \int_{u_{s}}^{u_{t}(b)}\frac{\diff u}{\sqrt{U(u, b)}}\\
&= - \left( \frac{\partial u_t}{\partial b} \frac{1}{\sqrt{U(u_t(b), b)}} - \int_{u_{o}}^{u_{t}(b)}\frac{\partial U / \partial b}{2 U^{3/2}}\diff u\right) - \left( \frac{\partial u_t}{\partial b} \frac{1}{\sqrt{U(u_t(b), b)}} - \int_{u_{s}}^{u_{t}(b)}\frac{\partial U / \partial b}{2 U^{3/2}}\diff u\right)\\
&= \int_{u_{o}}^{u_{t}(b)}\frac{\partial U / \partial b}{2 U^{3/2}}\diff u + \int_{u_{s}}^{u_{t}(b)}\frac{\partial U / \partial b}{2 U^{3/2}}\diff u - 2\frac{\partial u_t}{\partial b} \frac{1}{\sqrt{U(u_t(b), b)}}\ . 
\end{split}
\end{equation}

\newpage
\bibliography{references}

\end{document}